\documentclass[fleqn,usenatbib]{mnras}
\usepackage[T1]{fontenc}
\usepackage{ae,aecompl}
\usepackage{soul}

\usepackage[english]{babel}
\usepackage{graphicx}
\usepackage{amsmath,mathtools}
\usepackage{amssymb}
\usepackage{makecell}
\usepackage{bm}
\usepackage{csquotes}
\usepackage{xcolor}
\usepackage[%
linecolor=red,
innermargin=-5.75pt,
innerleftmargin=5pt,
outermargin=-5.75pt,
innerrightmargin=5pt,
innertopmargin=0pt,
innerbottommargin=0pt,
topline=false,
bottomline=false,
linewidth=0.75pt,
skipabove=0.4\baselineskip,
skipbelow=0.4\baselineskip,
]{mdframed}

\usepackage{aas_macros}

\colorlet{RED}{red}

\newcommand{\diff}[2]{{\frac{d{#1}}{d{#2}}}}
\newcommand{\pdiff}[2]{{\frac{\partial{#1}}{\partial{#2}}}}

\title[Turbulent convection in protoplanetary discs]{Turbulent convection in protoplanetary discs and its role in angular momentum transfer}

\author[E. P. Kurbatov and Ya. N. Pavlyuchenkov]{E. P. Kurbatov\thanks{E-mail: kurbatov@inasan.ru}
and Ya. N. Pavlyuchenkov
\\
$^{1}$Institute of Astronomy of the RAS, Moscow, Russia
}

\date{Accepted XXX. Received YYY; in original form ZZZ}

\pubyear{YYYY}

\begin{document}

\label{firstpage}
\pagerange{\pageref{firstpage}--\pageref{lastpage}}
\maketitle

\begin{abstract}
We present a model for the transport of anisotropic turbulence in an accretion disc. The model uses the Reynolds stress tensor approach in the mean field approximation. To study the role of convection in a protoplanetary disc, we combine the turbulence model with a radiative transfer calculation, and also include convection using the mixing length approximation. We find that the turbulence generated by convection causes the angular momentum of the accretion disc to be directed outwards. We also confirm the conclusions of other authors that turbulent convection is unable to provide the observed disc accretion rates as well as a heat source sufficient for the convection to be self-sustaining. The reasons for the latter are the strong anisotropy of the turbulence together with the low efficiency of the energy transfer from the background velocity shear to the turbulent stress tensor.
\end{abstract}


\section{Introduction}

The theory of disc accretion is used in astrophysics to explain a wide range of observed sources and phenomena: active galactic nuclei, the evolution and variability of close binary systems, the formation of jets and bipolar outflows, the structure of protoplanetary discs, the formation of planetary systems, and many others, see for example \cite{Shakura2018afa.book, Hartmann2009apsf.book.....H, Armitage2015arXiv150906382A}. In all these objects, accretion takes place under different physical conditions, varying in temperature, density, degree of ionisation, magnetic induction, radiation field, presence of dust, and so on. However, the processes that influence accretion have common consequences: this is the redistribution and removal of angular momentum, which allows matter to accrete from the disc to the central object.

Turbulence is thought to play an important role in many accretion processes. The classical approach to describe turbulent accretion relies on the formalism of turbulent viscosity, which is mathematically equivalent to molecular viscosity. In the \cite{Shakura1973A&A....24..337S} model, this formalism is reduced to the setting of an alpha parameter that relates the coefficient of turbulent viscosity to the speed of sound and the scale height of the disc. This phenomenological approach has proved extremely useful for describing the structure and evolution of astrophysical discs, but the question of the causes and properties of the turbulence itself remains beyond its scope.

Currently, one of the most active areas of astrophysical research is the study of protoplanetary discs (PPDs) around young stars. This interest is stimulated by the progress in observational techniques providing the means to obtain direct images of the discs at different wavelengths~\citep[see the review by][]{Andrews2020ARA&A..58..483A}. The high angular and spectral resolution allows, among other things, the reconstruction of the detailed distribution of the turbulent gas velocity across the disc, with the estimates varying greatly for different sources \citep[see][]{Flaherty2017ApJ...843..150F, Guilloteau2012A&A...548A..70G}. In this context, protoplanetary discs can be considered as a convenient natural laboratory for studying the physics of accretion and turbulence in general.

It is widely accepted that turbulence in the disc arises due to some instability. In protoplanetary discs, the possible triggers of turbulence could be gravitational, thermal, magneto-rotational, baroclinic, streaming, vertical shear and other instabilities \citep[see the reviews by][]{Armitage2015arXiv150906382A, Bae2022arXiv221013314B,Lesur2022arXiv220309821L}. These instabilities appear at different dynamical and thermodynamical conditions in the disc, \citep[see, e.g.,][]{Pfeil2019ApJ...871..150P}. Each instability is the subject of extensive research. For example, in \cite{Klahr2014ApJ...788...21K} the authors considered the effects of radial buoyancy in discs. It was shown that in a rotating flow, radial buoyancy together with centrifugal force can cause epicyclic oscillations with increasing amplitude \citep{Latter2016MNRAS.455.2608L, Volponi2016MNRAS.460..560V}, the phenomenon named convective overstability. In the nonlinear regime, this instability can lead to a subcritical baroclinic instability \citep{Lyra2014ApJ...789...77L}, as well as to the growth of large-scale vortices, which may play a role in planet formation \citep{Raettig2021ApJ...913...92R}.

In the present work we are interested in the convective instability. The link between convection, turbulence and angular momentum redistribution has been investigated in many studies. The idea that convection in protoplanetary discs can not only transfer heat but also provide viscosity and thus influence the evolution of the disc was formulated by \cite{Cameron1978M&P....18....5C} and \cite{Lin1980MNRAS.191...37L}. This idea has generated a lot of interest, but after several decades of research the role of convection in the transfer of angular momentum is still controversial, see a detailed historical review in \cite{Klahr2007IAUS..239..405K}, and also the recent papers by \cite{Held2018MNRAS.480.4797H, Held2021MNRAS.504.2940H}. A representative example is that in early numerical models, convection was found to cause the transfer of angular momentum towards the accretor \citep{Stone1996ApJ...464..364S}, which would correspond to a negative alpha parameter. In later work, using high-resolution numerical schemes, it was shown that the angular momentum of the accreting matter is transferred outwards \citep[see][and discussion therein]{Held2021MNRAS.504.2940H}. \cite{Held2018MNRAS.480.4797H} presented the results of 3D modelling of convection in a disc, illustrating the emergence of convective cells, eddies and other coherent structures upon the initiation of convection. At the same time, they noted that they could not obtain a self-sustaining convection regime in the disc. \cite{Held2021MNRAS.504.2940H} showed that the interaction of convective and magneto-rotational instabilities ensures the periodic nature of accretion in the disc. \cite{Pavlyuchenkov2020ARep...64....1P} and \cite{Maksimova2020ARep...64..815M} also showed that convective instability in a protoplanetary disc can lead to irregular accretion onto a star. This result is relevant in the context of the search for physical mechanisms to confirm the scenario of episodic accretion in protoplanetary discs \citep{Hartmann2009apsf.book.....H}, which is important for solving the problem of observed accretion luminosities and for explaining the nature of young stellar objects with luminosity outbursts, such as FU~Ori and EX~Lup type stars. However, the key approximation of the model presented in \cite{Pavlyuchenkov2020ARep...64....1P} is the assumption that the emerging convection is accompanied by high turbulent viscosity.

\begin{figure}
  \begin{center}
    \includegraphics[width=\columnwidth]{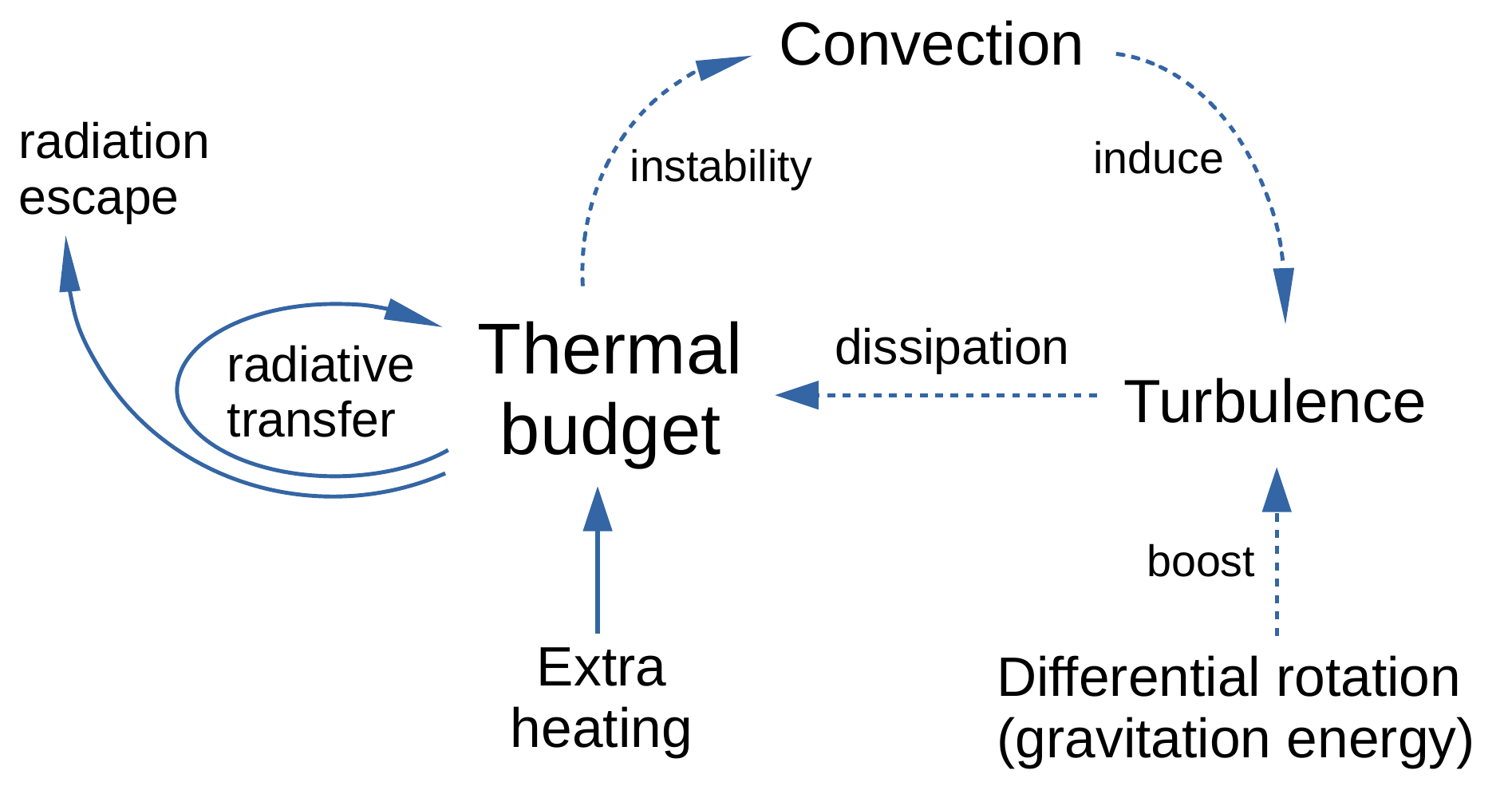}
  \end{center}
  \caption{Energy circulation in an accretion disc with a convective turbulence source.}
  \label{fig:scheme}
\end{figure}

The relationship between convection, turbulence and accretion in the disc layer is pictured in Fig.~\ref{fig:scheme}, which is based on the energy circulation scheme. Let us suppose that there is an initial heating source in the medium. Thermal energy is transferred by radiative diffusion and is eventually emitted as infrared radiation. Certain conditions can lead to the development of convection, which not only transfers some of the heat, but also excites the turbulence. The dissipation of the turbulence eventually converts the kinetic energy back into thermal energy. In addition to this cycle, the turbulence may be intensified by the background shear flow, which dissipates its kinetic energy and replenishes the heat budget, providing another energy source for convection. This way, the turbulence converts the gravitational energy of the gas into the thermal energy. In this scheme, the fundamental question is how significant is the contribution of the energy of the differential rotation to the turbulence strength.

Fig. \ref{fig:scheme} mentions a source of extra heating outside the energy cycle.
This source is necessary to trigger the convection, but its underlying physical mechanisms may vary. As an example, one can suggest the heating due to density waves excited by the disc self-gravity~\citep{2009MNRAS.393.1157C}. Another possible sources are the magneto-rotational instability~\citep{Held2018MNRAS.480.4797H} or dissipation of large scale magnetic field \citep{Bethune2020MNRAS.494.6103B}. Cosmic rays, which can penetrate quite deep into the disc, can also be considered as a source of heating \citep{DAlessio1998ApJ...500..411D}. In addition, there are factors that directly prevent convection. For example, the heating of the disc by stellar and interstellar radiation helps to establish a positive temperature gradient in the upper layers of the disc, making it stable against convection.

Although direct numerical simulations can produce very realistic results, a complete self-consistent three-dimensional calculation of the accretion disc evolution with a sufficiently high {spatial and temporal} resolution {remains a challenging task.} Even if the full-scale numerical model is {assumed to be} sufficiently resolved and accurate, {there remains the} problem of interpretation, i.e. the assessment of the importance of one or another physical factor that influences the gas dynamics, transport processes, etc. In this paper, we implement a non-isotropic turbulent transport model (in the mean-field approximation), together with a calculation of the convective flow (in the mixing-length approximation) and radiative transport to study the problem of turbulent convection and its role in the redistribution of angular momentum. The mean-field approach makes the calculations simpler than the full three-dimensional hydrodynamic calculations, while allowing the most important physical processes to be emphasized. {Thus we can explicitly determine} the contribution of different factors {involved in} the energy cycle to the dynamics of turbulence.

We  base our modelling on the mean field turbulence model proposed by \cite{Canuto1992ApJ...392..218C, Canuto1993ApJ...416..331C, Canuto1997ApJ...482..827C} for stellar atmospheres. This model relies on the momentum representation to describe the fields of velocity, density and pressure fluctuations up to the fourth order moments. It is quite complex and has never been fully implemented for astrophysical applications. We formulate a reduced version of the Canuto model, where only the dynamics of the turbulent stress tensor is calculated explicitly, while the remaining closures are implemented in the gradient or algebraic approximations. As in the original model of Canuto, the only seed of the turbulence is the convective flux. However, the turbulence can grow or decay due to interaction with the background shear flow. The second component of our model is infrared (IR) radiative transfer for PPDs by \cite{Pavlyuchenkov2020ARep...64....1P}. It uses temperature-dependent opacities for a mixture of graphite and silicate dust grains, which makes it possible to realistically simulate the conditions for the development of the convective instability. This model also takes into account the absorption of the radiation from the central star and the interstellar medium.

In Section~\ref{sec:meanfield_turb_model}, we present the mean-field turbulence transfer equations, as well as the closures and the convective flux. In Section~\ref{sec:app-onedim}, the full turbulent convection and the radiative transfer model is formulated in the one-dimensional cylindrical frame. We also perform the test calculations and implement the model for a vertical column in a protoplanetary disc. Discussion and conclusions are presented in Sections~\ref{sec:discussion} and~\ref{sec:conclusions}, respectively.

\section{Mean field turbulence model}
\label{sec:meanfield_turb_model}

\subsection{Mean field and turbulence transfer equations}

As mentioned in the introduction, in this paper we use the mean-field approach to model the turbulence. In this approach, we can distinguish two methods for the description of the turbulence: the filter method and the statistical method. The former uses a spatial and temporal filter, then the details of the sub-scale flow are only approximated by averaging on the filter scale. This approach is implemented in the class of subgrid models and in the Large Eddy Simulation \citep[LES,][]{Leonard1975AdGeo..18..237L, Meneveau1996JFM...319..353M}. Another approach is to introduce a statistical ensemble for the turbulent fluctuations, assuming that they are stochastic. The properties of the fluctuations are then formulated in terms of the statistical moments of this ensemble \citep[see e.g. the references in][]{Canuto1997ApJ...482..827C}. Since there is only one realisation of the flow in any given problem, the LES method may seem more physically justified. In addition, this method can explicitly describe non-local effects, such as the interaction of subgrid and supergrid scale structures \citep{Leonard1975AdGeo..18..237L, Stewart1976A&A....49...39S}. The statistical approach, in turn, greatly simplifies the computations, as it allows the use of empirical information on the amplitudes of the fluctuations and their mutual correlations. We are interested in the effect of the turbulence on the mean flow rather than in the detailed spatial and temporal structure of the turbulence, so we describe its properties in terms of the statistical moments of the velocity, density and pressure fluctuations.

We will only provide the final expressions for the turbulent transfer model of \cite{Canuto1992ApJ...392..218C, Canuto1997ApJ...482..827C} (see also references therein), without the detailed derivation. We write the equations in arbitrary curvilinear coordinates with the metric tensor $\varkappa^{ij}$. Later on, the model will be implemented in cylindrical frame. Let us define the variables that characterise turbulent flow: volume density $\rho$, velocity $v^i$, pressure $p$, internal energy $e$, and the Reynolds stress tensor $w_{ij}$. These quantities satisfy the dynamic equations:
\begin{gather}
  \label{eq:flow-density-mean}
  \pdiff{\rho}{t} + \nabla_k (\rho v^k) = 0
  \;,  \displaybreak[0]\\
  \label{eq:flow-momentum-mean}
  \pdiff{(\rho v^i)}{t} + \nabla_k (\rho v^i v^k)
  ={} - \nabla^i p - \nabla_k w^{ik} + \rho g^i
  \;,  \displaybreak[0]\\
  \label{eq:flow-energy-mean}
  \begin{multlined}
      \pdiff{(\rho e)}{t} + \nabla_k (\rho e v^k)  \\[-5pt]
      ={} - p \nabla_k v^k + q - \nabla_k F_\mathrm{conv}^k
      - \frac{B_k^k}{2} + \frac{\Pi_k^k}{2}
      + \epsilon  \;,
  \end{multlined}
  \displaybreak[0]\\
  \label{eq:flow-reynolds}
  \begin{multlined}
      \pdiff{w^{ij}}{t} + \nabla_k (w^{ij} v^k) + \nabla_k w^{ijk}  \\[-5pt]
      ={} - \bigl( w^{ik} \nabla_k v^j + w^{jk} \nabla_k v^i \bigr)
      + B^{ij} - \Pi^{ij} - \frac{2}{3}\,\epsilon \varkappa^{ij}  \;.
  \end{multlined}
\end{gather}
In addition to the quantities listed above, these equations also include gravitational acceleration $g^i$, heat source $q$, convective flow $F_\mathrm{conv}^i$, and several closures to the equation for the Reynolds stress tensor ($w^{ijk}$, $B^{ij}$, $\Pi^{ij}$, and $\epsilon$), which will be defined later.

Thermodynamical variables are related to each other via the ideal gas equation of state:
\begin{gather}
  \label{eq:thermodynamics-eos}
  p = \frac{\mathcal{R}}{\mu}\,\rho T  \;,\qquad
  e = c_\mathrm{v} T  \;,  \\
  \label{eq:thermodynamics-coeffs}
  c_\mathrm{p} = \frac{\gamma}{\gamma - 1}\,\frac{\mathcal{R}}{\mu}  \;,\qquad
  c_\mathrm{v} = \frac{1}{\gamma - 1}\,\frac{\mathcal{R}}{\mu}  \;,
\end{gather}
where $T$ is the temperature; $\mu$ is the weight of a gas particle in hydrogen atom mass units, $m_\mathrm{H}$; $\mathcal{R} = k_\mathrm{B}/m_\mathrm{H} = 8.25\times10^7$~erg$\:$g$^{-1}\:$K$^{-1}$ is the gas constant; $\gamma$ is the adiabatic index; $c_\mathrm{p}$ and $c_\mathrm{v}$ are the specific heat at constant pressure and volume, respectively.

\subsection{Closures}
\label{sec:closures}

The closures in the r.h.s. of the Eqs.~\eqref{eq:flow-energy-mean} and \eqref{eq:flow-reynolds} are 2nd and 3rd-order statistical moments of turbulent fluctuations. It is also possible to formulate dynamic equations for these moments. However, since the model already contains a large number of parameters, we will use algebraic closures. Making the model more complex will only make the results more difficult to interpret.

A valuable quantity in the algebraic closures is the turbulence correlation time. Under accretion disc conditions, Keplerian time appears to be a natural time scale for turbulence correlation. An estimate obtained by \citet{Stewart1976A&A....49...39S} by analysing an equation similar to Eq.~\eqref{eq:flow-momentum-mean} leads to an expression for the correlation time of the form
\begin{equation}
  \label{eq:turb-corr-time}
  t_\mathrm{T}
  = \left( 1 + \mathcal{M}_\mathrm{T}^{-2} \right)^{1/2} |\Omega|^{-1}  \;,
\end{equation}
where $|\Omega|$ is the angular velocity of the gas rotation; $\mathcal{M}_\mathrm{T}$ is a turbulent Mach number, it depends on the local speed of sound $c_\mathrm{s}^2$:
\begin{equation}
  \label{eq:turb-mach-number}
  \mathcal{M}_\mathrm{T}^2
  = \frac{w_k^k}{\rho c_\mathrm{s}^2}  \;.
\end{equation}
Mach numbers estimated from the non-thermal broadening of spectral lines in protoplanetary discs range from $0.06$ \citep{Flaherty2017ApJ...843..150F} to $0.5$ \citep{Guilloteau2012A&A...548A..70G}, which gives estimates for turbulence correlation time within $0.3 \lesssim |\Omega|\,t_\mathrm{T}/(2\pi) \lesssim 2.7$.

The $B^{ij}$ tensor is responsible for buoyancy effects. It can be expressed in terms of the convective flux:
\begin{equation}
  \label{eq:tensor-buoyancy}
  B^{ij}
  ={} - \frac{1}{c_\mathrm{p} T}
    \left( \frac{\nabla^i p}{\rho}\,F_\mathrm{conv}^j
    + \frac{\nabla^j p}{\rho}\,F_\mathrm{conv}^i \right)  \;.
\end{equation}
The buoyancy tensor is the seed of turbulence in the present model. Note that the r.h.s of Eq.~\eqref{eq:flow-energy-mean} contain the convective source $B_k^k$. In an accretion disc, one can expect the pressure gradient to be directed towards the disc mid-plane, while the convective flow is directed away from it. Hence we can conclude that $B_k^k \geqslant 0$, which means that convection takes away the thermal energy of the mean flow.

It is widely accepted that pressure fluctuations play a key role in the development of turbulence. In the $\Pi^{ij}$ tensor these effects manifest themselves in isotropising the turbulence (\enquote{return-to-isotropy}), buoyancy and interacting with the background flow. The following form of this tensor was derived from symmetry and dimension considerations \citep[see references in][]{Launder1974CMAME...3..269L,Speziale1991AnRFM..23..107S,Canuto1997ApJ...482..827C}:
\begin{multline}
  \label{eq:tensor-ret-to-isotropy}
  \Pi^{ij}
  = \frac{\mathbb{C}_{\Pi 1}}{t_\mathrm{T}}\,b^{ij}
    - \mathbb{C}_{\Pi 2} \left( b^{ik} U_k^j + b^{jk} U_k^i - \frac{2}{3}\,b^{kl} U_{kl} \varkappa^{ij} \right)  \\
    \hspace{0.2cm} - \mathbb{C}_{\Pi 3} \varkappa_{kl}\,\bigl( b^{ik} V^{jl} + b^{jk} V^{il} \bigr)
    - \left( U^{ij} - \frac{U_l^l}{3}\,\varkappa^{ij} \right) \frac{2w_k^k}{5}  \\
    + (1 - \mathbb{C}_B)\,B^{ij}  \;,
\end{multline}
where $b^{ij}$ characterises the deviation from isotropy,
\begin{equation}
  b^{ij}
  = w^{ij} - \frac{w_k^k}{3}\,\varkappa^{ij}  \;.
\end{equation}
$U^{ij}$ and $V^{ij}$ are the symmetric and asymmetric parts of the strain rate tensor of the background flow, respectively,
\begin{gather}
  \label{eq:strain_rate_symm}
  U^{ij}
  = \frac{1}{2} \left( \nabla^j v^i + \nabla^i v^j \right)  \;,  \\
  \label{eq:strain_rate_asymm}
  V^{ij}
  = \frac{1}{2} \left( \nabla^j v^i - \nabla^i v^j \right)  \;.
\end{gather}
Note that the expression \eqref{eq:tensor-ret-to-isotropy} depends linearly on the components of the Reynolds tensor, except for the last term.

The variable $\epsilon$ is related to the dissipation of turbulent energy. This process takes place on small scales where the anisotropy is on average weak. For this reason, we use the classical isotropic closure
\begin{equation}
  \label{eq:scalar-dissipation-rate}
  \epsilon
  = \frac{K}{t_\mathrm{T}}  \;,
\end{equation}
and $K = (1/2)\,w_k^k$ is the turbulent kinetic energy volume density.

To write out a closed expression for the third-order velocity momentum $w^{ijk}$, we use the gradient approximation:
\begin{equation}
  \label{eq:3rd-vel-corr-grad-approx}
  w^{ijk}
  = {}- \nu_\mathrm{T}\,\bigl( \nabla^i w^{jk} + \nabla^j w^{ki} + \nabla^k w^{ij} \bigr)  \;.
\end{equation}
The factor $\nu_\mathrm{T}$ is interpreted as the turbulent kinematic viscosity coefficient. It can be written as follows
\begin{equation}
  \label{eq:turb-visc-coeff}
  \nu_\mathrm{T}
  = \mathbb{C}_\nu t_\mathrm{T}\,\frac{K}{\rho}  \;.
\end{equation}
The dimensionless factor $\mathbb{C}_\nu$ is usually set equal to $0.09$ \citep{Launder1974CMAME...3..269L, Speziale1991AnRFM..23..107S}. As can be seen, the tensor $w^{ijk}$ describes the diffusion of turbulence.

The constants in the expressions \eqref{eq:tensor-ret-to-isotropy} and \eqref{eq:turb-visc-coeff} were obtained experimentally \citep[see papers by][and references therein]{Launder1974CMAME...3..269L,Speziale1991AnRFM..23..107S,Canuto1992ApJ...392..218C, Canuto1993ApJ...416..331C}:
\begin{equation}
  \label{eq:constants}
  \begin{aligned}
    & \mathbb{C}_\nu = 0.09  \;,\quad
      \mathbb{C}_B = 0.6  \;,  \\
    & \mathbb{C}_{\Pi 1} = 3.5  \;,\quad
      \mathbb{C}_{\Pi 2} = 0.61  \;,\quad
      \mathbb{C}_{\Pi 3} = 0.44  \;.
  \end{aligned}
\end{equation}

Finally, let  us consider the gas energy balance. Taking the trace of Eq.~\eqref{eq:flow-reynolds}, one can get
\begin{multline}
  \label{eq:flow-turb-energy}
  \pdiff{K}{t}
    + \nabla_k\!\left( K v^k + \frac{1}{2} \varkappa_{ij} w^{ijk} \right)  \\
  ={} - w^{jk} \nabla_k v_j + \frac{B_k^k}{2} - \frac{\Pi_k^k}{2} - \epsilon  \;.
\end{multline}
Here one can see that the $B_k^k$ source enters the r.h.s. with a positive sign. By the comparison with Eq.~\eqref{eq:flow-energy-mean}, we can conclude that convection converts the thermal energy of the mean flow into the turbulent energy.

Projecting Eq.~\eqref{eq:flow-momentum-mean} onto the velocity vector gives the equation for the kinetic energy of the mean flow:
\begin{multline}
  \label{eq:flow-kin-energy-mean}
  \pdiff{}{t}\!\left( \frac{\rho |v|^2}{2} \right)
    + \nabla_k\!\left( \frac{\rho |v|^2}{2}\,v^k \right)  \\
  ={} - v^k \nabla_k p - v_j \nabla_k w^{jk} + \rho v_j g^j  \;.
\end{multline}
Combining Eqs.~\eqref{eq:flow-energy-mean}, \eqref{eq:flow-turb-energy} and \eqref{eq:flow-kin-energy-mean}, we get the equation for the total energy:
\begin{multline}
  \label{eq:flow-total-energy-mean}
  \pdiff{}{t}\!\left( \frac{\rho |v|^2}{2} + K + \rho e \right)  \\
  \hspace{1cm}+ \nabla_k\!\left[ \left( \frac{\rho |v|^2}{2} + K + \rho e + p \right) v^k
    + w^{jk} v_j \right.  \\
    \left.{} + \frac{1}{2}\,\varkappa_{ij} w^{ijk}
    + F_\mathrm{conv}^k \right]
  = \rho v_j g^j + q  \;.
\end{multline}
This equation is completely conservative except for external sources. This means that the thermal, kinetic and turbulent energy of the medium can only transform to each other.

\subsection{Convective flux and conditions for instability}

In Mixing Length Theory (MLT), heat is transferred by convective elements formed by convective instability. It is usually assumed that the elements move with a characteristic velocity $v_\mathrm{conv}$ under the effect of the buoyancy force and that they transfer the excess heat $\rho c_\mathrm{p} \Delta T$ to the surroundings. The convective flux can be written as $\rho c_\mathrm{p} \Delta T v_\mathrm{conv}$. Different versions of the theory differ in the way the quantities $\Delta T$ and $v_\mathrm{conv}$ are estimated. In this way, the radiative heat losses of the convective element along its path, its viscous deceleration, and the spreading of convection beyond the convective zone (overshooting) \citep{Canuto1992ApJ...392..218C} can be considered.

The conditions for convective instability are fulfilled in the regions where the temperature gradient exceeds the adiabatic (more precisely, the isentropic) gradient in the direction opposite to the gravitational acceleration. Let us denote the excess temperature gradient as
\begin{equation}
  \label{eq:temperature-gradient-excess}
  \beta^i
  ={} - \left[ n^i n_k \nabla^k T - (\nabla^i T)_\mathrm{ad} \right]  \;,
\end{equation}
where $n^i$ is the unit vector in the direction of the gravitational acceleration $g^i$ (note that the centrifugal acceleration also contributes to the value of $g^i$). The adiabatic gradient is also expressed in terms of the acceleration vector,
\begin{equation}
  (\nabla^i T)_\mathrm{ad}
  = \frac{g^i}{c_\mathrm{p}}  \;.
\end{equation}
The instability condition is $- g_k \beta^k > 0$.

Let us write down the results given in \cite{Hansen1994sipp.book.....H} for the problem of stellar convection, without going into a detailed derivation. The characteristic parameter of the theory is the length of the mixing path $\ell$. This is the distance traveled by the convective element before it mixes with the surrounding matter. Another characteristic parameter is the growth rate of the convective instability \citep[Chap.~5]{Hansen1994sipp.book.....H},
\begin{equation}
  \label{eq:convective-increment}
  \omega
  ={} - \frac{\nu_\mathrm{mol} + \nu_\mathrm{rad}}{2\ell^2}
    + \left[ \frac{(\nu_\mathrm{mol} + \nu_\mathrm{rad})^2}{4\ell^4} + |\mathcal{N}|^2 \right]^{1/2}  \;,
\end{equation}
where $\nu_\mathrm{mol}$ and $\nu_\mathrm{rad}$ are the molecular (collisional) and radiative thermometric conductivities, respectively, \st{cm$^2/$s} cm$^2\:$s$^{-1}$; $\mathcal{N}$ is the Brunt-V{\"a}is{\"a}l{\"a} frequency, it is defined as
\begin{equation}
  \mathcal{N}^2
  = \frac{g_k \beta^k}{T}  \;.
\end{equation}
In a convectively stable medium ($g_k \beta^k > 0$), $\mathcal{N}$ is the frequency at which the gas element oscillates due to buoyancy and gravity. In the limit of weak convection, $|\mathcal{N}| \ll (\nu_\mathrm{mol} + \nu_\mathrm{rad})/(2\ell^2)$, the increment is $\omega \approx \ell^2 |\mathcal{N}|^2 / (\nu_\mathrm{mol} + \nu_\mathrm{rad})$. In the opposite limit the increment is saturated as $\omega \approx |\mathcal{N}|$.

The collisional thermometric conductivity coefficient for neutral atoms is
\begin{equation}
  \nu_\mathrm{mol}
  = \frac{1}{\rho c_\mathrm{p}}\,\frac{\mu m_\mathrm{H} c_\mathrm{v} v_\mathrm{th}}{3 \sigma_\mathrm{nn}}  \;,
\end{equation}
where $v_\mathrm{th} = (3 \mathcal{R} T/\mu)^{1/2}$ is the mean thermal velocity of the molecules; $\sigma_\mathrm{nn} = 3\times10^{-16}$ cm$^2$ is the col\-li\-sion cross section for neutral hydrogen. The radiative thermometric conductivity coefficient has the form
\begin{equation}
  \nu_\mathrm{rad}
  = \frac{1}{\rho c_\mathrm{p}}\,\frac{4c a_\mathrm{rad} T^3}{3\rho \kappa_\mathrm{R}}  \;,
\end{equation}
here $c$ is the speed of light; $a_\mathrm{rad} = 7.56\times10^{-15}$ erg$\:$cm$^{-3}\:$K$^{-4}$ is the radiation density constant; $\kappa_\mathrm{R}$ is the Rosseland mean opacity. It should be noted that in many astrophysical applications, the collisional mechanism of the heat conduction can be neglected.

The speed of the convective element and the excess temperature are estimated as follows:
\begin{gather}
  \label{eq:convective-velocity}
  v_\mathrm{conv}^i = \omega \ell n^i
  \;,  \\
  \Delta T
  = \frac{\omega^2}{|\mathcal{N}|^2}\,\ell\,|\beta|  \;,
\end{gather}
where $|\beta|$ is the magnitude of the vector $\beta^i$, Eq.~\eqref{eq:temperature-gradient-excess}. As a result, the convective energy flux takes the form
\begin{equation}
  \label{eq:convective-flux-mlt}
  F_\mathrm{conv}^i
  = \rho c_\mathrm{p}\,\frac{\omega^3}{|\mathcal{N}|^2}\,\ell^2 \beta^i  \;.
\end{equation}

The MLT has one free parameter, the mixing path length $\ell$. The convective flux \eqref{eq:convective-flux-mlt} is very sensitive to this parameter: from $\ell^8$ in weak convection to $\ell^2$ in strong convection. The pressure scale height of a star is usually adopted as the mixing length in stellar convective shells models. Under the accretion disc conditions, the thermal scale height of the disc can be taken as the mixing length.

\section{Model of turbulent convection in a protoplanetary disc}
\label{sec:app-onedim}

\subsection{Final system of equations}

Consider a cylindrical coordinate system\footnote{%
Further, all vector and tensor quantities will be written in local Cartesian projections.
}
$(r, \phi, z)$ and a rotational axisymmetric flow in a narrow radial annulus of radius $r$. Our requirement is that all quantities are independent of the azimuthal angle. It will be assumed that the disc is close to a mechanical equilibrium. In this case, the gradients can be estimated as
\begin{equation}
  \pdiff{}{r} \lesssim \pdiff{}{z} \sim \frac{1}{H} \sim \frac{|\Omega|}{c_\mathrm{s}}  \;,
\end{equation}
where $H$ is the vertical thermal scale in the disc; $\Omega = v_\phi/r$ is the angular velocity of the gas rotation. If no luminosity outbursts are considered in the disc, then the radial and vertical velocities of the gas in typical accreting discs become essentially subsonic:
\begin{gather}
  |v_r| \sim |v_z| \equiv \mathcal{M} c_\mathrm{s}
  \;,  \\
  \left| \pdiff{v_r}{r} \right| \sim \left| \pdiff{v_r}{z} \right|
  \sim \left| \pdiff{v_z}{r} \right| \sim \left| \pdiff{v_z}{z} \right|
  \sim \frac{\mathcal{M} c_\mathrm{s}}{H} \sim \mathcal{M} |\Omega|  \;.
\end{gather}
where $\mathcal{M} \ll 1$ is the Mach number for the radial and vertical background gas velocities. In this approximation the advection terms in the expressions \eqref{eq:flow-momentum-mean}--\eqref{eq:flow-reynolds} can be neglected, except for the centrifugal force in the Euler equation, as well as the corresponding components in the turbulence transport equation. For all quantities except the background gas angular velocity, the radial dependence is neglected. We also neglect the self-gravity of the disc and consider only the gravity of the star. Finally, the system of equations \eqref{eq:flow-momentum-mean}--\eqref{eq:flow-reynolds} takes the following form:
\begin{gather}
  \label{eq:flow-momentum-cyl}
  \pdiff{p}{z}
  = \rho \Omega_\mathrm{K}^2 z - \pdiff{w_{zz}}{z}  \;,
  \displaybreak[0]\\
  \label{eq:flow-energy-cyl}
  \rho c_\mathrm{v}\,\pdiff{T}{t}
  = - \pdiff{F_\mathrm{conv}}{z} - \frac{\mathbb{C}_B}{2}\,B_{zz} + \epsilon + q  \;,
  \\
  \label{eq:flow-reynolds-cyl-wrr}
  \pdiff{w_{rr}}{t}
  = \pdiff{}{z} \left( \nu_\mathrm{T}\,\pdiff{w_{rr}}{z} \right)
  - \Pi_{rr} + 4\Omega w_{r\phi}
  - \frac{2\epsilon}{3}  \;,
  \\
  \label{eq:flow-reynolds-cyl-wff}
  \pdiff{w_{\phi\phi}}{t}
  = \pdiff{}{z} \left( \nu_\mathrm{T}\,\pdiff{w_{\phi\phi}}{z} \right)
  - \Pi_{\phi\phi} - \frac{2}{r} \pdiff{(r^2\Omega)}{r}\,w_{r\phi}
  - \frac{2\epsilon}{3}  \;,
  \\
  \label{eq:flow-reynolds-cyl-wzz}
  \pdiff{w_{zz}}{t}
  = \pdiff{}{z} \left( 3\nu_\mathrm{T}\,\pdiff{w_{zz}}{z} \right)
  - \Pi_{zz} + B_{zz}
  - \frac{2\epsilon}{3}  \;,
  \\
  \label{eq:flow-reynolds-cyl-wrf}
  \pdiff{w_{r\phi}}{t}
  = \pdiff{}{z} \left( \nu_\mathrm{T}\,\pdiff{w_{r\phi}}{z} \right)
  - \Pi_{r\phi} - \frac{1}{r} \pdiff{(r^2\Omega)}{r}\,w_{rr} + 2\Omega w_{\phi\phi}  \;,
  \displaybreak[0]\\
  \label{eq:flow-reynolds-cyl-wrz}
  \pdiff{w_{rz}}{t}
  = \pdiff{}{z} \left( 2\nu_\mathrm{T}\,\pdiff{w_{rz}}{z} \right)
  - \Pi_{rz} + 2\Omega w_{\phi z}  \;,
  \displaybreak[0]\\
  \label{eq:flow-reynolds-cyl-wfz}
  \pdiff{w_{\phi z}}{t}
  = \pdiff{}{z} \left( 2\nu_\mathrm{T}\,\pdiff{w_{\phi z}}{z} \right)
  - \Pi_{\phi z} - \frac{1}{r} \pdiff{(r^2\Omega)}{r}\,w_{rz} - \Omega w_{\phi z}  \;.
\end{gather}
In Eq. \eqref{eq:flow-momentum-cyl}, $\Omega_\mathrm{K}$ is the Keplerian angular velocity at the radius $r$; $\nu_\mathrm{T}$ is the turbulent viscosity coefficient \eqref{eq:turb-visc-coeff}.

Due to the chosen approximation, only the $z$-component of the convective flux vector \eqref{eq:convective-flux-mlt} remains non-zero providing the only non-zero component of the buoyancy tensor:
\begin{equation}
  B_{zz}
  = - \frac{2}{C_\mathrm{p} \rho T}\,\pdiff{p}{z}\,F_\mathrm{conv}
  \;.
\end{equation}
The components of the isotropisation tensor are now
\begin{gather}
  \label{eq:Pi_rr}
  \Pi_{rr}
  = \frac{\mathbb{C}_\mathrm{\Pi 1}}{t_\mathrm{T}}\,b_{rr}
  - \left( \frac{2\mathbb{C}_\mathrm{\Pi 2}}{3}\,U_{r\phi}
  - 2 \mathbb{C}_\mathrm{\Pi 3} V_{r\phi} \right) b_{r\phi}
  \;,  \displaybreak[0]\\
  \Pi_{\phi\phi}
  = \frac{\mathbb{C}_\mathrm{\Pi 1}}{t_\mathrm{T}}\,b_{\phi\phi}
  - \left( \frac{2\mathbb{C}_\mathrm{\Pi 2}}{3}\,U_{r\phi}
  + 2 \mathbb{C}_\mathrm{\Pi 3} V_{r\phi} \right) b_{r\phi}
  \;,  \displaybreak[0]\\
  \Pi_{zz}
  = \frac{\mathbb{C}_\mathrm{\Pi 1}}{t_\mathrm{T}}\,b_{zz}
  + \frac{4\mathbb{C}_\mathrm{\Pi 2}}{3}\,U_{r\phi} b_{r\phi}
  + (1 - \mathbb{C}_B)\,B_{zz}  \;,  \displaybreak[0]\\
  \label{eq:Pi_rf}
  \begin{multlined}[b]
      \Pi_{r\phi}
      = \frac{\mathbb{C}_\mathrm{\Pi 1}}{t_\mathrm{T}}\,b_{r\phi}
      - \mathbb{C}_\mathrm{\Pi 2} U_{r\phi} \left( b_{rr} + b_{\phi\phi} \right)  \\[-3pt]
      \qquad\qquad - \mathbb{C}_\mathrm{\Pi 3} V_{r\phi} \left( b_{rr} - b_{\phi\phi} \right)
      - \frac{4}{5}\,U_{r\phi} K  \;,
  \end{multlined}
  \displaybreak[0]\\
  \Pi_{rz}
  = \frac{\mathbb{C}_\mathrm{\Pi 1}}{t_\mathrm{T}}\,b_{rz}
  - \left( \mathbb{C}_\mathrm{\Pi 2} U_{r\phi}
  - \mathbb{C}_\mathrm{\Pi 3} V_{r\phi} \right) b_{\phi z}
  \;,  \displaybreak[0]\\
  \Pi_{\phi z}
  = \frac{\mathbb{C}_\mathrm{\Pi 1}}{t_\mathrm{T}}\,b_{\phi z}
  - \left( \mathbb{C}_\mathrm{\Pi 2} U_{r\phi}
  + \mathbb{C}_\mathrm{\Pi 3} V_{r\phi} \right) b_{rz}
  \;,
\end{gather}
where
\begin{gather}
  b_{rr} = \frac{1}{3}\,(2w_{rr} - w_{\phi\phi} - w_{zz})  \;,  \displaybreak[0]\\
  b_{\phi\phi} = \frac{1}{3}\,(2w_{\phi\phi} - w_{rr} - w_{zz})  \;,  \displaybreak[0]\\
  b_{zz} = \frac{1}{3}\,(2w_{zz} - w_{rr} - w_{\phi\phi})  \;,  \displaybreak[0]\\
  b_{r\phi} = w_{r\phi}  \;,\qquad
    b_{rz} = w_{rz}  \;,\qquad
    b_{\phi z} = w_{\phi z}  \;,  \displaybreak[0]\\
  K = \frac{1}{2}\,(w_{rr} + w_{\phi\phi} + w_{zz})  \;,  \displaybreak[0]\\
  U_{r\phi} = \frac{r}{2}\,\pdiff{\Omega}{r}  \;,\qquad
    V_{r\phi} = - \frac{1}{2r}\,\pdiff{(r^2 \Omega)}{r}  \;.
\end{gather}

As can be seen from the expressions \eqref{eq:flow-reynolds-cyl-wrr}--\eqref{eq:flow-reynolds-cyl-wfz}, the convective heat flux is the seed of turbulence. The background flow is only involved in the amplification or weakening of different $w_{ij}$ components. For the MLT flux \eqref{eq:convective-flux-mlt} we take the mixing length to be equal to the vertical thermal scale of the disc, $\ell \equiv H = c_\mathrm{s}/|\Omega|$.

There are several sources on the r.h.s. of the heat balance equation \eqref{eq:flow-energy-cyl}: the first one is responsible for convective heat transfer, the second one describes the consumption of thermal energy for convective motions, the third one provides energy input due to turbulence dissipation. The last source, $q$, accounts for heating by stellar and interstellar radiation, exchanging energy with its own IR radiation, and may also include an additional heat source. The model for $q$ is given in Section \ref{sec:radtrans}.

The Eqs. \eqref{eq:flow-reynolds-cyl-wrr}--\eqref{eq:flow-reynolds-cyl-wfz} require boundary conditions. At the upper boundary of the disc, it is natural to set each of $w_{ij}$ to zero. In the disc mid-plane ($z = 0$), the boundary conditions are as follows:
\begin{gather}
  \label{eq:reynolds-boundary-1}
  \pdiff{w_{rr}}{z} = \pdiff{w_{\phi\phi}}{z} = \pdiff{w_{zz}}{z} = \pdiff{w_{r\phi}}{z} = 0
  \;,  \\
  \label{eq:reynolds-boundary-2}
  w_{rz} = w_{\phi z} = 0  \;.
\end{gather}

It can be seen that in the system \eqref{eq:flow-reynolds-cyl-wrr}--\eqref{eq:flow-reynolds-cyl-wfz}, the last two equations do not contain any energy source or sink and only govern the redistribution of the turbulent energy between the components $w_{rz}$ and $w_{\phi z}$. Given the boundary conditions, this means that if these components are initially zero, they will remain zero in the future. Thus, the equations for the $w_{rz}$ and $w_{\phi z}$ are not considered further.

\subsection{Testing the model of turbulent transfer}
\label{sec:testing}

Here we will briefly analyse the turbulent transfer model and try to reveal the role of the free parameters $\mathbb{C}_{\Pi 1}$--$\mathbb{C}_{\Pi 3}$ defined in Eq.~\eqref{eq:constants}. From Eqs.~\eqref{eq:Pi_rr}--\eqref{eq:Pi_rf} one can see that $\mathbb{C}_{\Pi 1}/t_\mathrm{T}$ is the inverse characteristic time of Reynolds stress tensor isotropisation. The parameters $\mathbb{C}_{\Pi 2}$ and $\mathbb{C}_{\Pi 3}$, in turn, are the coupling constants between the turbulence and the background shear flow (its symmetric and asymmetric parts, respectively, see Eqs.~\eqref{eq:strain_rate_symm} and \eqref{eq:strain_rate_asymm}). Finally, $\mathbb{C}_B B_{zz}/2$ is the heat loss per unit time due to the excitation of turbulence by the convection channel.

Let us convert the field variables to the dimensionless form:
\begin{gather}
  w_{ij} \mapsto \tilde{w}_{ij} = \frac{w_{ij}}{\rho c_\mathrm{s}^2}  \;,
  \\
  t \mapsto \tau = |\Omega|\,t  \;,
  \\
  t_\mathrm{T} \mapsto \tau_\mathrm{T}
  = |\Omega|\,t_\mathrm{T}
  = \frac{(1 + \mathcal{M}_\mathrm{T}^2)^{1/2}}{\mathcal{M}_\mathrm{T}}  \;,
  \\
  \mathcal{M}_\mathrm{T}^2 = 2 \tilde{K}  \;,\qquad
  \tilde{K} \equiv \frac{1}{2}\,(\tilde{w}_{rr} + \tilde{w}_{\phi\phi} + \tilde{w}_{zz})  \;,
  \\
  B_{zz} \mapsto \tilde{B}_{zz} \equiv \frac{B_{zz}}{|\Omega|\,\rho c_\mathrm{s}^2}  \;.
\end{gather}

We simplify the model by neglecting turbulent diffusion, which is applicable to the conditions deep within the convective zone. In this case the equations \eqref{eq:flow-reynolds-cyl-wrr}--\eqref{eq:flow-reynolds-cyl-wrf} are reduced to a system of ODEs. The local background shear velocity profile is assumed to be in general form, rather than Keplerian, $\Omega \propto r^{-q}$. It is important to note that $\Omega$ is the projection of the angular velocity vector onto the $OZ$ axis. It can be shown that the $w_{r\phi}$ component always enters the equations in a combination $\operatorname{sign}(\Omega)\,w_{r\phi}$. Thus, the inversion of the angular velocity leads to a change of the sign of $w_{r\phi}$ (and also of $w_{rz}$ and $w_{\phi z}$, see Eqs.~\eqref{eq:flow-reynolds-cyl-wrf}--\eqref{eq:flow-reynolds-cyl-wfz} and the corresponding equations for $\Pi_{ij}$). Without loss of generality we assume $\Omega > 0$.

It is useful to regroup the dimensionless equations to the following representation:
\begin{gather}
  \label{eq:dimensionless_wrr+wff}
  \diff{(\tilde{w}_{rr} + \tilde{w}_{\phi\phi})}{\tau} ={}
  \begin{multlined}[t]
    - \frac{\mathbb{C}_{\Pi 1}}{\tau_\mathrm{T}}\,(\tilde{w}_{rr} + \tilde{w}_{\phi\phi})
    + \frac{\mathbb{C}_{\Pi 1} - 1}{3 \tau_\mathrm{T}}\,4 \tilde{K}  \\
    + \left( 1 - \frac{\mathbb{C}_{\Pi 2}}{3} \right) 2 q \tilde{w}_{r\phi}  \;,
  \end{multlined}
  \\
  \label{eq:dimensionless_wrr-wff}
  \diff{(\tilde{w}_{rr} - \tilde{w}_{\phi\phi})}{\tau} ={}
  \begin{multlined}[t]
    - \frac{\mathbb{C}_{\Pi 1}}{\tau_\mathrm{T}}\,(\tilde{w}_{rr} - \tilde{w}_{\phi\phi})  \\
    + \bigl[ \mathbb{C}_{\Pi 3}\,(2 - q) + 4 - q \bigr] 2 \tilde{w}_{r\phi}  \;,
  \end{multlined}
  \\
  \label{eq:dimensionless_urf}
  \diff{\tilde{w}_{r\phi}}{\tau} ={}
  \begin{multlined}[t]
    - \frac{\mathbb{C}_{\Pi 1}}{\tau_\mathrm{T}}\,\tilde{w}_{r\phi}
    + \left( \frac{\mathbb{C}_{\Pi 2}}{3} - \frac{1}{5} \right) 2 q \tilde{K}  \\
    \hspace{-0.75cm}- \frac{\mathbb{C}_{\Pi 2}}{2}\,q\,(\tilde{w}_{rr} + \tilde{w}_{\phi\phi})
    - \frac{\mathbb{C}_{\Pi 3}}{2}\,(2 - q)\,(\tilde{w}_{rr} - \tilde{w}_{\phi\phi})  \;,
  \end{multlined}
  \\
  \label{eq:dimensionless_k}
  \diff{\tilde{K}}{\tau} ={}
  - \frac{\tilde{K}}{\tau_\mathrm{T}} + q \tilde{w}_{r\phi}
  + \frac{\mathbb{C}_B}{2}\,\tilde{B}_{zz}  \;.
\end{gather}
It is seen here that the component $\tilde{w}_{r\phi}$ is coupled to $\tilde{w}_{rr} + \tilde{w}_{\phi\phi}$ by $\mathbb{C}_{\Pi 2}$ and to $\tilde{w}_{rr} - \tilde{w}_{\phi\phi}$ by $\mathbb{C}_{\Pi 3}$.

In a steady state limit, the energy equation gives
\begin{equation}
  \label{eq:steady-state-limit}
  q \tilde{w}_{r\phi}
  = \frac{\tilde{K}}{\tau_\mathrm{T}} - \frac{\mathbb{C}_B}{2}\,\tilde{B}_{zz}  \;.
\end{equation}
The $\tilde{w}_{r\phi}$ component is therefore determined by the difference between the amount of {dissipated} turbulent energy and the energy supplied by the convective source. Despite that convection is the seed of energy for turbulence in our model, the background velocity shear is also important. Once the convection {is} able to {produce} the turbulence, the velocity shear starts to amplify the components of the Reynolds stress tensor (through the coupling constants $\mathbb{C}_{\Pi 2}$ and $\mathbb{C}_{\Pi 3}$), making the r.h.s. of the Eq.~\eqref{eq:steady-state-limit} non-zero. Since the $\tilde{w}_{r\phi}$ component is responsible for the angular momentum transfer in the disc, the efficiency of the transfer is clearly related to the energy balance. When there is no background shear, $q = 0$, the parameter $\tilde{w}_{r\phi}$ is decoupled from the turbulent energy $\tilde{K}$, so there is no transfer of angular momentum.

Let us assume $\tilde{w}_{rr} = \tilde{w}_{\phi\phi}$. Then it can be shown from Eqs.~\eqref{eq:dimensionless_wrr+wff}--\eqref{eq:dimensionless_urf} that in the steady-state limit all the turbulence components are zero. Since this derivation is independent of the source of the turbulence (in r.h.s. of the Eq.~\eqref{eq:dimensionless_k}), this is true not only for convection-generated turbulence, but also in the general case of rotational shear flows. The latter means that turbulence is always anisotropic in accretion discs (at least in the no-diffusion approximation).

\begin{figure}
  \begin{center}
    \includegraphics[width=\columnwidth]{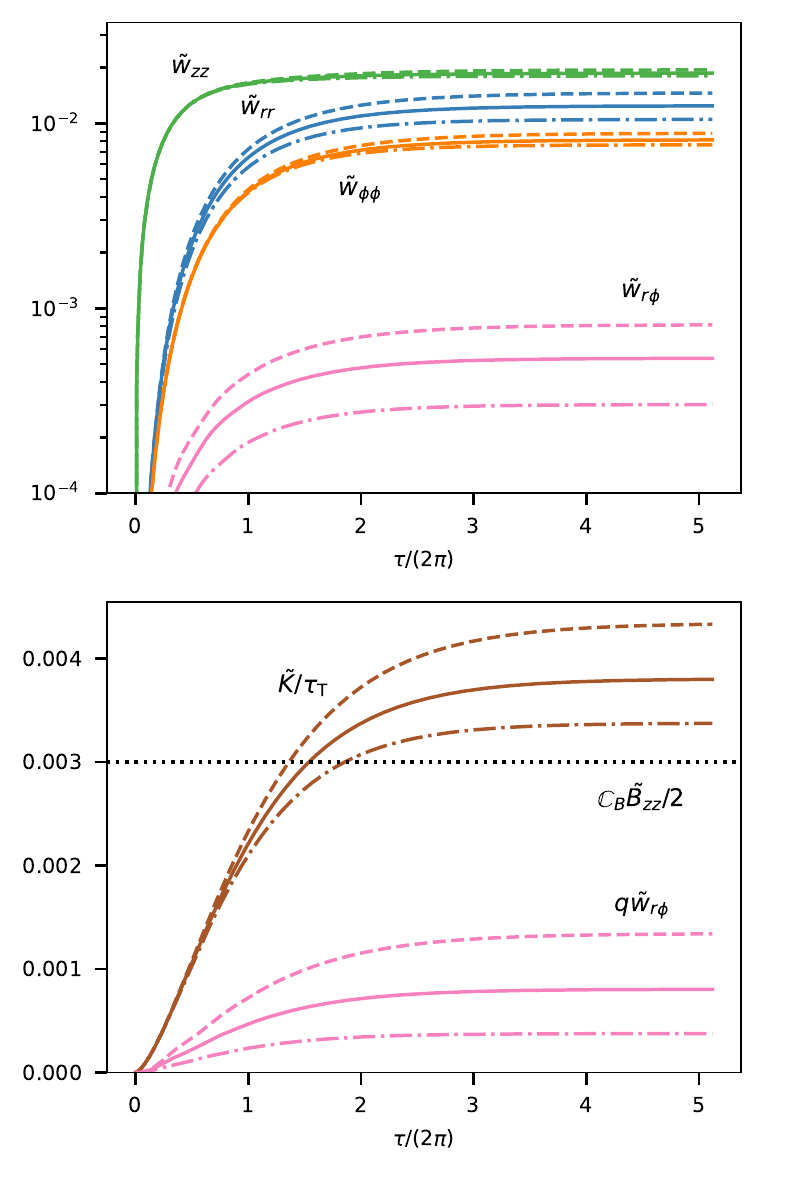}
  \end{center}
  \caption{Components of the Reynolds stress tensor and related quantities in the dimensionless local model (Sec.~\ref{sec:testing}). {Different values of the velocity profile indices were explored}: $q = 1.5$ (\textit{solid curves}), $q = 1.65$ (\textit{dashed}), $q = 1.25$ (\textit{dot-dashed}).}
  \label{fig:probe}
\end{figure}

We have performed calculations of the model \eqref{eq:dimensionless_wrr+wff}--\eqref{eq:dimensionless_k} with a fixed convective source $\tilde{B}_{zz} = 0.01$ and fixed values of the constants $\mathbb{C}_B$, $\mathbb{C}_{\Pi 1}$, $\mathbb{C}_{\Pi 2}$ and $\mathbb{C}_{\Pi 3}$ from the Eq.~\eqref{eq:constants}. Fig.~\ref{fig:probe} shows that after a monotonic growth phase lasting $1$--$3$ disc periods, the turbulence reaches a steady state. Varying the velocity profile index $q$ by $10\%$--$15\%$ significantly affects the turbulence intensity, in particular the $\tilde{w}_{r\phi}$ component changes by a factor of two. Simulations with high velocity profile indices revealed that $q \tilde{w}_{r\phi} \sim \tilde{K}/\tau_\mathrm{T} \sim \tilde{K}$ for $q \gtrsim 3$ (not shown in Fig.~\ref{fig:probe}). Negative values of $q$ leads to the negative values of $\tilde{w}_{r\phi}$, though the dependence on $q$ is weaker.

\begin{figure*}
  \begin{center}
    \includegraphics[width=\textwidth]{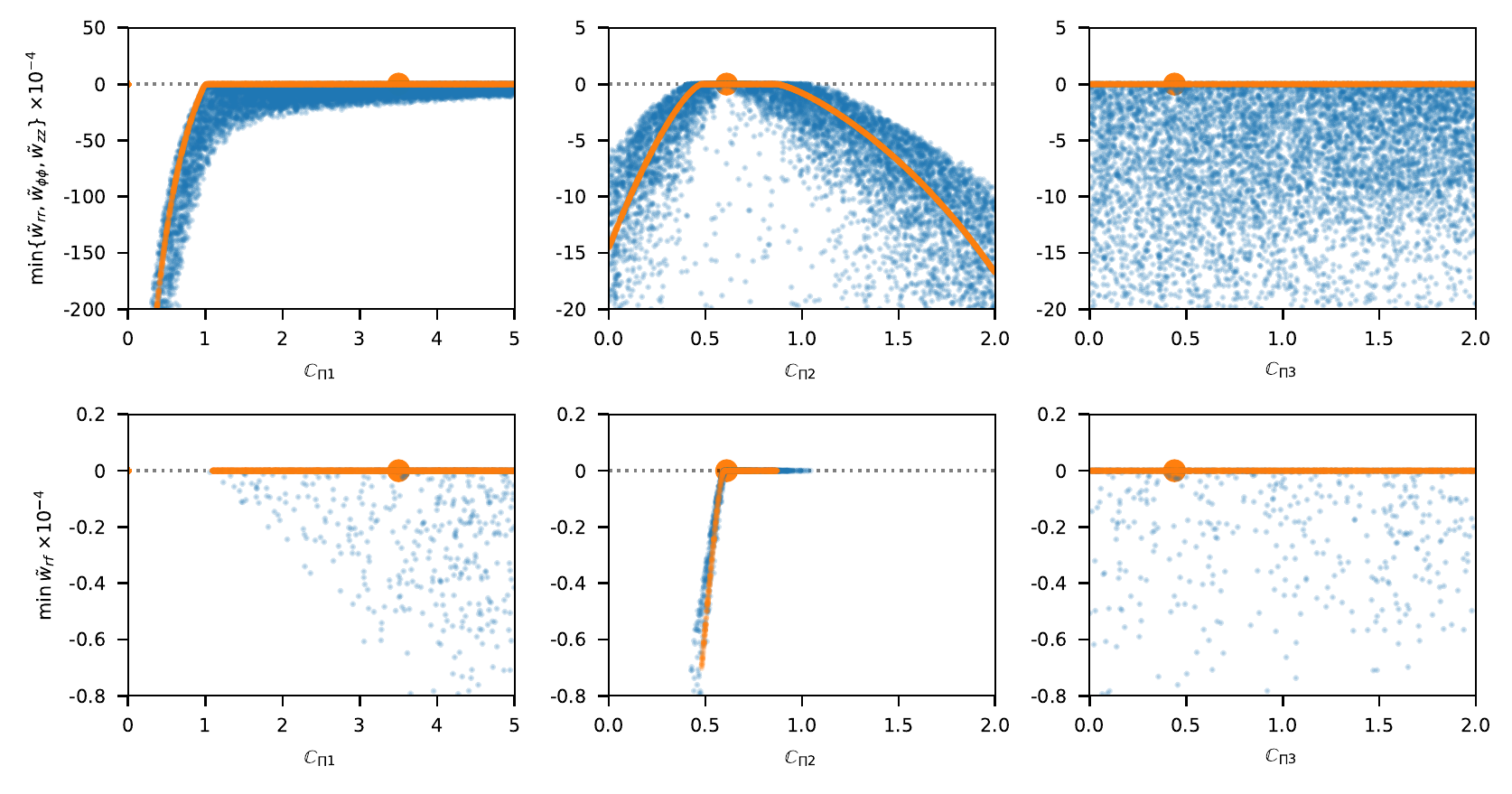}
  \end{center}
  \caption{Qualitative indicators of the solutions to system \eqref{eq:dimensionless_wrr+wff}--\eqref{eq:dimensionless_k}, for $q = 1.5$. \textit{Top row:} Minimum values of the components of the Reynolds tensor diagonal over the whole simulation time. \textit{Bottom row:} Minimum values of the off-diagonal component $\tilde{w}_{r\phi}$ over the simulation time. Each dot is a simulation for some set of the constant parameters. The pictures in the bottom row show only the points corresponding to solutions with a non-negative diagonal, i.e. physically allowed solutions. The set of blue dots is obtained by {uniformly} sampling all three parameters: $0 \leq \mathbb{C}_{\Pi 1} \leq 5$, $0 \leq \mathbb{C}_{\Pi 2} \leq 2$, $0 \leq \mathbb{C}_{\Pi 3} \leq 2$. The orange dots are obtained by sampling only the constant parameter that labels the horizontal axis, while the other two parameters are fixed at their experimental values \eqref{eq:constants}. The latter are also marked with the thick dots.}
  \label{fig:random_samples}
\end{figure*}

\begin{figure}
  \begin{center}
    \includegraphics[width=\columnwidth]{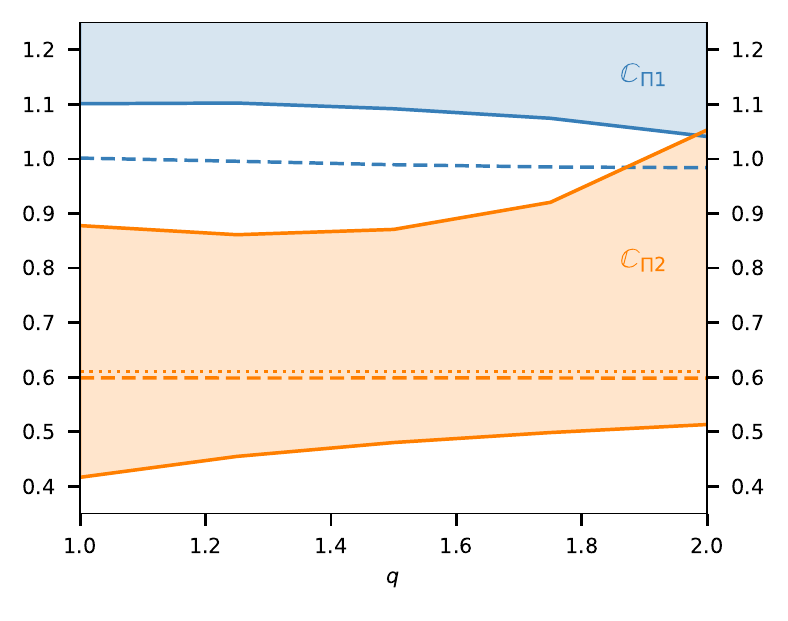}
  \end{center}
  \caption{\textit{Solid lines:} Limits of the allowed values of the $\mathbb{C}_{\Pi 1}$ (blue shaded area) and $\mathbb{C}_{\Pi 2}$ (orange shaded area) {parameters} according to physical constraint $\min\{\tilde{w}_{rr}, \tilde{w}_{\phi\phi}, \tilde{w}_{zz}\} \geq 0$, depending on the velocity profile index. \textit{Dashed lines:} Lower limits for $\mathbb{C}_{\Pi 1}$ (blue) and $\mathbb{C}_{\Pi 2}$ (orange), according to the condition $\min \tilde{w}_{r\phi} \geq 0$. The dotted orange line is the experimental value of $\mathbb{C}_{\Pi 2}$.}
  \label{fig:param_bounds}
\end{figure}

It is interesting to see how the choice of constant parameter values affects turbulence. For random sets of the constant parameters and various velocity profile indices, we ran a ensemble of test calculations, starting from zero initial conditions. The convective source was fixed to $\tilde{B}_{zz} = 0.01$ as it only determines the value of the turbulence energy in the steady state limit, see Eq.~\eqref{eq:steady-state-limit}. The constants were chosen randomly from the intervals $0 \leq \mathbb{C}_{\Pi 1} \leq 5$, $0 \leq \mathbb{C}_{\Pi 2} \leq 2$, and $0 \leq \mathbb{C}_{\Pi 3} \leq 2$ (blue dots in Fig.~\ref{fig:random_samples}). The velocity profile index $q$ was set to $1.5$. Two qualitative indicators of the solutions are shown in Fig.~\ref{fig:random_samples}. The first indicator, $\min\{\tilde{w}_{rr}, \tilde{w}_{\phi\phi}, \tilde{w}_{zz}\}$ declares physical constraints: the quadratic velocity correlators should not be negative. Solutions that satisfy this constraint leave this indicator at zero value. It is seen that the physically allowed values of $\mathbb{C}_{\Pi 1}$ cannot be lower than $\sim 1$. The allowed values of $\mathbb{C}_{\Pi 2}$ are bounded in a quite narrow range around the experimental value \eqref{eq:constants}. Varying $\mathbb{C}_{\Pi 3}$ within the considered limits does not violate the physical constraints. The second indicator in Fig.~\ref{fig:random_samples}, $\min \tilde{w}_{rf}$, shows under which conditions the direction of the angular momentum flux changes. Note that only the physically allowed solutions are shown here, i.e. the solutions with $\min\{\tilde{w}_{rr}, \tilde{w}_{\phi\phi}, \tilde{w}_{zz}\} = 0$. The sign of the off-diagonal component $\tilde{w}_{r\phi}$ is not sensitive to $\mathbb{C}_{\Pi 1}$ and $\mathbb{C}_{\Pi 3}$, however, it is sensitive to $\mathbb{C}_{\Pi 2}$. Surprisingly, the experimental value of $\mathbb{C}_{\Pi 2}$ from Eq.~\eqref{eq:constants} only slightly exceeds the lower limit of the range where $\min \tilde{w}_{r\phi} \geq 0$. The above boundaries have been estimated from the considered ensemble runs and are shown in more detail in Fig.~\ref{fig:param_bounds}.

\begin{figure*}
  \begin{center}
    \includegraphics[width=\textwidth]{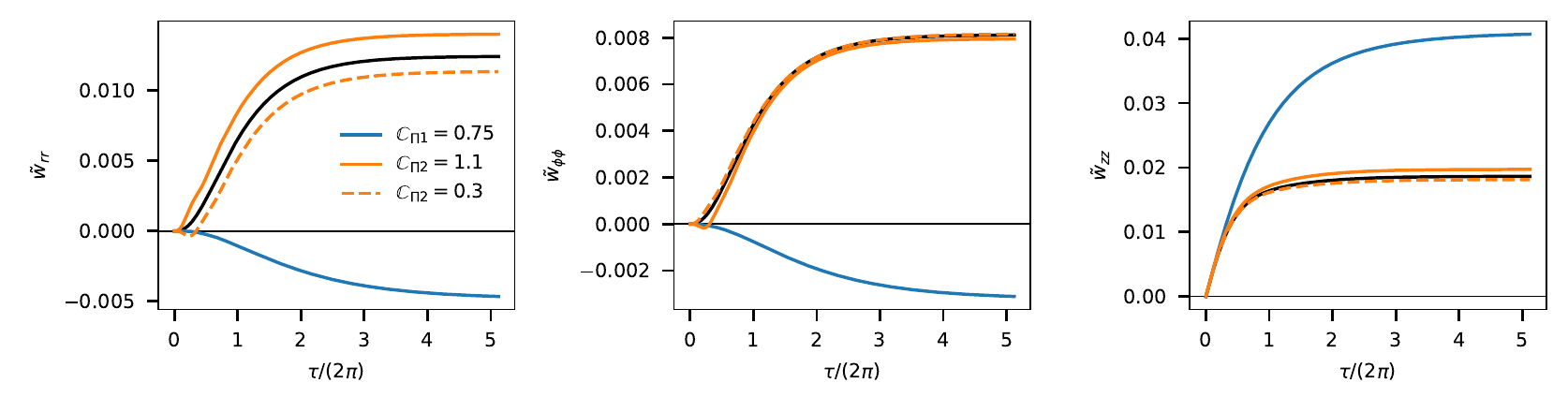}
  \end{center}
  \vspace{-0.5cm}
  \begin{center}
    \includegraphics[width=\textwidth]{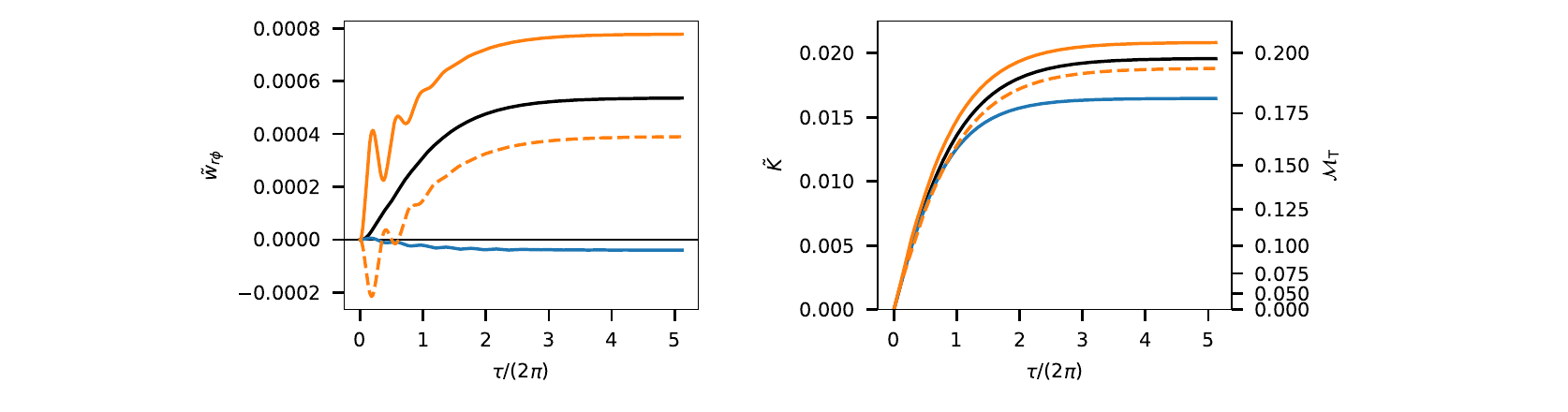}
  \end{center}
  \caption{Local turbulent convection model with $q = 1.5$, where different constant parameters are tested. \textit{Black lines:} The experimental values of the constants, see Eq.~\eqref{eq:constants}. \textit{Coloured lines:} One of the constants is varied (see the legend {in} the first plot).}
  \label{fig:probe_vary}
\end{figure*}

Fig.~\ref{fig:probe_vary} shows the effect of the outlying values of the constant parameters on the qualitative behavior of the model. Assigning $\mathbb{C}_{\Pi 1}$ a value below the acceptable range leads to the negative steady-state limit in $\tilde{w}_{rr}$ and $\tilde{w}_{\phi\phi}$. On the other hand, the assignment of $\mathbb{C}_{\Pi 2}$ to the outlying value results in oscillating (including negative) solutions but positive limits. Interestingly, in these calculations the turbulence energy is always positive and quite stable.

\subsection{Combining the turbulent convection model with the radiative transfer model}
\label{sec:radtrans}

We calculate radiative transfer and heat balance using the thermal model of \cite{Vorobyov2017A&A...606A...5V} and \cite{Pavlyuchenkov2020ARep...64....1P}. The model considers the energy exchange between the gas and the IR radiation, the UV heating by stellar and interstellar radiation ($S_\mathrm{UV}$), and an additional heat source ($S_\mathrm{ext}$):
\begin{equation}
  \label{eq:heat-source-tot}
  q
  = c \rho \kappa_\mathrm{P}\,(E_\mathrm{rad} - a_\mathrm{rad} T^4)
    + \rho S_\mathrm{UV} + \rho S_\mathrm{ext}  \;,
\end{equation}
where $E_\mathrm{rad}$ is the energy volume density of the IR radiation; $\kappa_\mathrm{P}$ is the Planck mean absorption coefficient. The $S_\mathrm{ext}$ source can be associated with some dissipation processes that are not explicitly included in our model (see below in this section). The heating due to the stellar and interstellar irradiation, $S_\mathrm{UV}$, is calculated by a direct integration of the radiative transfer \citep{Vorobyov2017A&A...606A...5V}. The IR radiative transfer is simulated by means of the Eddington approximation, which is reduced to the following system of equations:
\begin{gather}
  \label{eq:radtrans}
  \pdiff{E_\mathrm{rad}}{t} + \pdiff{F_\mathrm{rad}}{z}
  ={} - c \rho \kappa_\mathrm{P}\,(E_\mathrm{rad} - a_\mathrm{rad} T^4)
  \;,  \\
  \label{eq:flux-rad}
  F_\mathrm{rad}
  ={} - \frac{c}{3 \rho \kappa_\mathrm{R}}\,\pdiff{E_\mathrm{rad}}{z}  \;,
\end{gather}
where $F_\mathrm{rad}$ is the IR radiation flux; $\kappa_\mathrm{R}$ is the Rosseland mean absorption coefficient. The radiation flux boundary conditions in the mid-plane and at the upper boundary $z_\mathrm{max}$ are of the form
\begin{gather}
  F_\mathrm{rad} \bigr|_{z=0} = 0
  \;,  \\
  \pdiff{F_\mathrm{rad}}{z} \biggr|_{z=z_\mathrm{max}}
  = \frac{c}{2} \left( E_\mathrm{rad} \Bigr|_{z=z_\mathrm{max}} - a_\mathrm{rad} T_\mathrm{CMB}^4 \right)  \;,
\end{gather}
where $T_\mathrm{CMB} = 2.73$~K. An important feature of the disc thermal model is the use of temperature dependent Rosseland and Planck mean opacities, as it has been found that an increase in opacity with temperature is necessary for the onset of convection \citep{Lin1980MNRAS.191...37L}. The opacities have been taken from \cite{Pavlyuchenkov2020ARep...64....1P} where they were obtained for a mixture of graphite and silicate dust grains. The solution of the subsystem of the radiative transfer equations \eqref{eq:heat-source-tot}--\eqref{eq:flux-rad} is found by an implicit method. The details and tests are described in the appendix of \cite{Vorobyov2017A&A...606A...5V}.

\begin{figure*}
  \begin{center}
    \includegraphics[width=\textwidth]{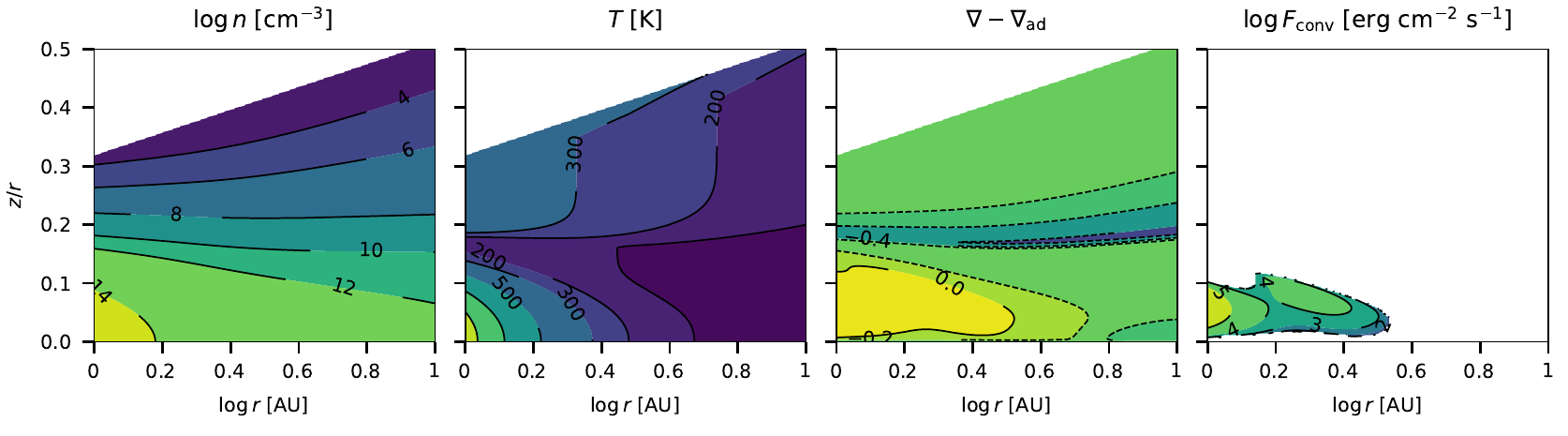}
  \end{center}
  \caption{The structure of a protoplanetary disc with no convective energy transport and no turbulence. From left to right: gas number density, temperature, temperature gradient excess, and the magnitude of the convective flux.}
  \label{fig:2d-pane}
\end{figure*}

We noted above that in our model, convection is the seed of turbulence. To initiate convection, an external heat source is required. It is included in the heat energy equation as $S_\mathrm{ext}$ in Eq.~\eqref{eq:heat-source-tot}. We define the external energy injection rate by using the accretion rate $\dot{M}$ as a measure \citep{Shakura1973A&A....24..337S}:
\begin{equation}
  \label{eq:external-heating}
  S_\mathrm{ext}
  = \left| \diff{\ln \Omega}{\ln r} \right| \frac{\dot{M} \Omega^2}{4\pi \Sigma}  \;,
\end{equation}
where $\Sigma$ is the local surface density of the disc.

Figure~\ref{fig:2d-pane} shows the thermal structure of a protoplanetary disc obtained in the 1+1-dimensional framework using the radiation transfer model without convection and turbulence. By 1+1 formalism, we mean that self-consistent density and temperature distributions in vertical direction (the first `1'-dimension) are obtained separately for each radial position (the second `1'-dimension ) of the disk. In our calculations, the surface density distribution and stellar heating depend on the radial position, but the columns do not affect each other. The purpose of this calculation was to locate the strongest convective instability. This model includes all the heating sources from the one-dimensional model above, along with the external heating $S_\mathrm{ext}$. The parameters of the model are: the stellar mass $1$~$M_\odot$, the star effective temperature $5780$~K, the accretion rate onto the star $10^{-7}$ $M_\odot\:$yr$^{-1}$, the density $10^3$~g$\:$cm$^{-2}$ at a radius of $1$~au, assuming a power law of the surface density $\Sigma\propto r^{-1}$. The convective flow was included in this calculation, but was not involved in the formation of the disc thermal structure. The last two plots in the Fig.~\ref{fig:2d-pane} show the region of convective instability ($\nabla - \nabla_\mathrm{ad} > 0$%
\footnote{%
This is the excess temperature gradient with respect to the pressure: $\nabla - \nabla_\mathrm{ad} \equiv dT/dp - (dT/dp)_\mathrm{ad}$. This gradient is calculated along the direction of the buoyancy force (in our case along the $OZ$ axis).}).
The convection is concentrated in a rather shallow torus near the mid-plane (not reaching it due to symmetry constraints). The maximum value of the convective flux is reached in the inner part of the computational domain, $r \approx 1$~au, and drops by three orders of magnitude at $r \approx 3$~au. Note that the flux distribution has two peaks, \citep[cf. with][Fig.~10]{Pfeil2019ApJ...871..150P}.

We have performed several runs considering different radial positions of the disc column. We have chosen the column at a radius of $3.34$~au for two reasons. First, as the average temperature of the disc decreases with distance from the star, the role of radiative heat transfer within the convective zone is reduced in favour of convective heat transfer. Second, the thickness of the convective zone decreases with radial distance. This choice allows us to see the effect of turbulent diffusion in detail.

\subsection{Numerical solution scheme}
\label{sec:numerics}

The solution of the full system of equations \eqref{eq:flow-momentum-cyl}--\eqref{eq:flow-reynolds-cyl-wfz} for each time step was divided in two stages. In the first stage, the heat balance, radiative transfer and hydrostatic equations were solved jointly with the given sources $S_\mathrm{UV}$, $S_\mathrm{ext}$ and $w_{ij}$. Within a time step, the heat balance equation was linearised in temperature and the hydrostatic equation was linearised in density. In the radiative transfer equation, the spatial derivative operator was discretised using a standard scheme and expressed by a tridiagonal matrix. This system of equations was solved using a completely implicit iterative scheme. The detailed scheme of the solution is described in \cite{Vorobyov2017A&A...606A...5V} and \cite{Pavlyuchenkov2020ARep...64....1P}.

In the second stage, the turbulence transfer equations were solved. This was done by discretisation the spatial derivative operator and solving the entire system of equations on the spatial grid as a system of ODEs in time. The solution was performed using an explicit-implicit scheme LSODA.\footnote{%
The model was implemented on \textsc{Python~3.7} \citep{vanRossum2009doi10.5555/1593511} + \textsc{Numpy} \citep{Harris2020array} + \textsc{Scipy}  \citep{2020SciPy-NMeth} + \textsc{Numba} \citep{Lam2015numba} + \textsc{Numbalsoda} \citep{Wogan2022numbalsoda}. The solution of the turbulence transfer equation was carried out according to an explicit-implicit scheme using the LSODA algorithm \citep{Petzold1983doi:10.1137/0904010}.}

\subsection{Disc column with turbulent convection}
\label{sec:column}

The vertical structure of the disc column was modelled in the one-dimensional turbulent disc convection approach. We set the values of the external heat source $S_\mathrm{ext}$ by {parameterising} it with the accretion rate, Eq.~\eqref{eq:external-heating}: $\dot{M} = 10^{-7}$ $M_\odot\:$yr$^{-1}$ (Model A) and $\dot{M} = 10^{-4}$ $M_\odot\:$yr$^{-1}$ (Model B). These options correspond to a quiescent and {outburst} state of the disc, respectively \citep{Audard2014prpl.conf..387A, Fischer2022arXiv220311257F}.

Previously, we ran a 1+1-dimensional disc model in order to find the {location} of the convective zone and to select the most interesting conditions for a full calculations including turbulent convection. However, that model did not take the convective heat transfer into account. As a result, the density and temperature distributions were unstable and therefore not physical. {As will be seen later, the account of the convective heat transfer significantly changes the steady state of the disc. We start the full model with the same initial conditions as in the previous model, namely assuming an isothermal disc with a temperature of $100$~K.} The main parameters of the model are listed in Table \ref{tbl:input}. Calculations were continued until steady states were reached. In total, Model A was run for $127$ Keplerian orbits, and the Model B was run for eight orbits. We also run each model without convection (hence, without turbulence) {to have a baseline to compare our convective models to}. The model without convection consists of the hydrostatic \eqref{eq:flow-momentum-cyl}, heat transfer \eqref{eq:flow-energy-cyl}, and radiative transfer \eqref{eq:heat-source-tot}--\eqref{eq:flux-rad} equations, only: all the convection and turbulence terms ($w_{zz}$, $F_\mathrm{conv}$, $B_{zz}$, and $\epsilon$) have been omitted in this model.

Using the proposed model, it is {interesting} to evaluate how effectively {the} convective turbulence utilizes external heating and background shear flow. It is also important to investigate whether the energy cycle (Fig.~\ref{fig:scheme}) can be self-sustaining without external heating, leading to steady convection and turbulence. The {purpose} of this modelling is to quantify the effective accretion rate and the dissipation rate of turbulence.

\begin{table}
  \centering
  \caption{Parameters of disc models with turbulent convection}
  \begin{tabular}{lrl}
    \hline  \\[-3mm]
    Stellar mass & $M_\mathrm{s}$ & $1~M_\odot$  \\
    Stellar radius & $R_\mathrm{s}$ & $1~R_\odot$  \\
    Stellar effective temperature & $T_\mathrm{s}$ & $5780$ K  \\
    Radial distance of the column & $r$ & $3.34$ au  \\
    Gas molecular weight & $\mu$ & $2.3$  \\
    Gas adiabatic exponent & $\gamma$ & $7/5$  \\
    Disc surface density & $\Sigma$ & $542$ g$\:$cm$^{-2}$  \\
    Gas number density  \\
      \qquad\quad at the external boundary & $n_\mathrm{ext}$ & $10^3$ cm$^{-3}$  \\
    ISM radiation temperature & $T_\mathrm{ISR}$ & $10^4$ K  \\
    \\
    Accretion rate (Model A) & $\dot{M}$ & $10^{-7}$ $M_\odot\:$yr$^{-1}$  \\
    \hphantom{Accretion rate} (Model B) & $\dot{M}$ & $10^{-4}$ $M_\odot\:$yr$^{-1}$  \\
    \hline
  \end{tabular}
  \label{tbl:input}
\end{table}

\subsubsection{Model A ($\dot{M} = 10^{-7}$ $M_\odot\:$yr$^{-1}$)}

\begin{figure*}
  \begin{center}
    \includegraphics[width=\textwidth]{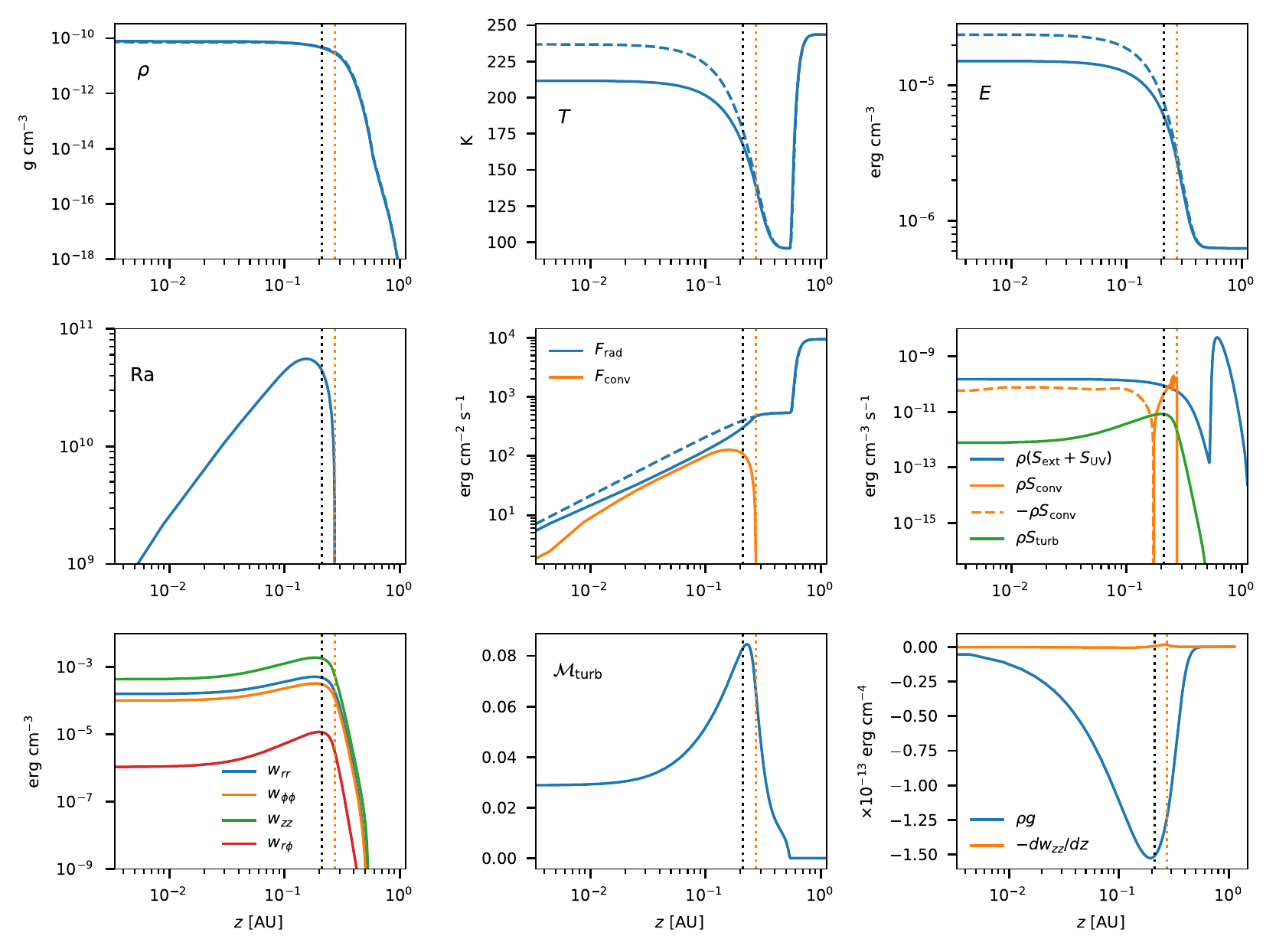}
  \end{center}
  \caption{The Model A results, $\dot{M} = 10^{-7}$ $M_\odot\:$yr$^{-1}$. Top row, from left to right: density, temperature, bulk radiation density. Middle row: Rayleigh number, energy flux, volumetric heat source (yellow dashed line denotes negative $\rho S_\mathrm{conv}$ values). Bottom row: turbulent stress tensor components, turbulent Mach number and volumetric force density. The blue dashed lines indicate the results of the model without taking into account convection. Dotted orange vertical lines mark the top boundary of the convective zone. Dotted black vertical lines mark the thermal scale of the disc.}
  \label{fig:model-a}
\end{figure*}

The results for Model A are shown in Fig. \ref{fig:model-a}. As can be seen, the convective zone extends upward to approximately one and a half thermal scale heights. In a significant part of the convective zone, the convective flow turns out to be comparable in magnitude to the radiative flow. Thus, convection provides about half of the total heat flow. The total energy flux is only a few percent higher than the energy flux in the calculation without convection (dashed lines). This, however, is enough to reduce the temperature of the inner layers of the disc, within the thermal scale, by about $25$ K.

In our model, the convective flow is directed along the $OZ$ axis, so the convection only excites the $w_{zz}$ component of the Reynolds tensor directly. The remaining components of this tensor arise from $w_{zz}$ due to the pressure tensor $\Pi_{ij}$, then amplify through interaction with the background flow. Note that $w_{zz}$ dominates in magnitude, it exceeds $w_{rr}$ and $w_{\phi\phi}$ by $3$--$5$ times and exceeds $w_{r\phi}$ by two or more orders of magnitude. The intensity of turbulence and the corresponding contribution to the width of the observed spectral lines is determined by the diagonal sum of the Reynolds tensor, $\sum_k w_{kk}$. Due to the turbulent diffusion, the region of developed turbulence is almost twice as thick as the convective zone. In the bulk of the disc material, the turbulent Mach number $\mathcal{M}_\mathrm{T}$ varies in the range $0.03$--$0.08$, which is in good agreement with the estimates obtained from observations \citep{Flaherty2017ApJ...843..150F}.

The rate of angular momentum removal due to the turbulence and the corresponding accretion rate (which we will call the effective accretion rate) depend on the off-diagonal component $w_{r\phi}$
\citep{Shakura1973A&A....24..337S}:
\begin{equation}
  \label{eq:accr-rate-eff}
  \dot{M}_\mathrm{eff}
  = \frac{2\pi}{|\Omega|}\,W_{r\phi}  \;,
\end{equation}
\begin{equation}
  W_{r\phi}
  = 2 \int_0^{z_\mathrm{max}} w_{r\phi}\, dz \;.
\end{equation}
The effective accretion rate can also be estimated in terms of the alpha parameter:
\begin{gather}
 \alpha_\mathrm{eff}
  = \frac{W_{r\phi}}{P}  \;,
  \\
 P
  = 2 \int_0^{z_\mathrm{max}} \frac{\mathcal{R}}{\mu}\,\rho T \,dz.
\end{gather}
due to the fact that $w_{r\phi} \ll \sum_k w_{kk}$ (see Fig.~\ref{fig:model-a} and \st{the} Table \ref{tbl:output}), one cannot draw a correct conclusion about the accretion rate from the turbulent kinetic energy, or, equivalently, from the width of the spectral lines. In the Model A, the effective accretion rate is $\dot{M}_\mathrm{eff} = 1.7\times10^{-10}$ $M_\odot\:$yr$^{-1}$, which is $1000$ times less than the accretion rate $\dot{M}$ that determines the heating of the layer.

Fig. \ref{fig:model-a} (middle row, right column) shows the distribution of external heating sources $\rho (S_\mathrm{ext} + S_\mathrm{UV})$, as well as the source associated with convection $\rho S_\mathrm{conv}$ and with turbulent dissipation, $\rho S_\mathrm{turb}$:
\begin{gather}
 \rho S_\mathrm{conv}
  = - \pdiff{F_\mathrm{conv}}{z} - \frac{\mathbb{C}_B}{2}\,B_{zz}  \;,
  \\
 \rho S_\mathrm{turb}
  = \epsilon  \;.
\end{gather}
In the inner layers of the disc, $z < 0.5$ au, the external source $\rho S_\mathrm{ext}$ dominates among all sources, except for a narrow region near the top boundary of the convective zone, where the convective source $\rho S_\mathrm{conv}$ is more important. Heating from the convective source is comparable to external heating in absolute value, but its role is to redistribute and slightly reduce thermal energy. The source $\rho S_\mathrm{turb}$ associated with turbulence dissipation is smaller than $\rho S_\mathrm{ext}$ by one or two orders of magnitude everywhere in the disc. The integrated values of each of the sources ($Q = 2 \int_0^{z_\mathrm{max}} \rho S\,dz$) are given in Table~\ref{tbl:output}. Also note that in the Model A even more energy is spent to start convection than is returned to the heat budget from dissipation: $|Q_\mathrm{conv}| > Q_\mathrm{turb}$.

\subsubsection{Model B ($\dot{M} = 10^{-4}$ $M_\odot\:$yr$^{-1}$)}

\begin{figure*}
  \begin{center}
    \includegraphics[width=\textwidth]{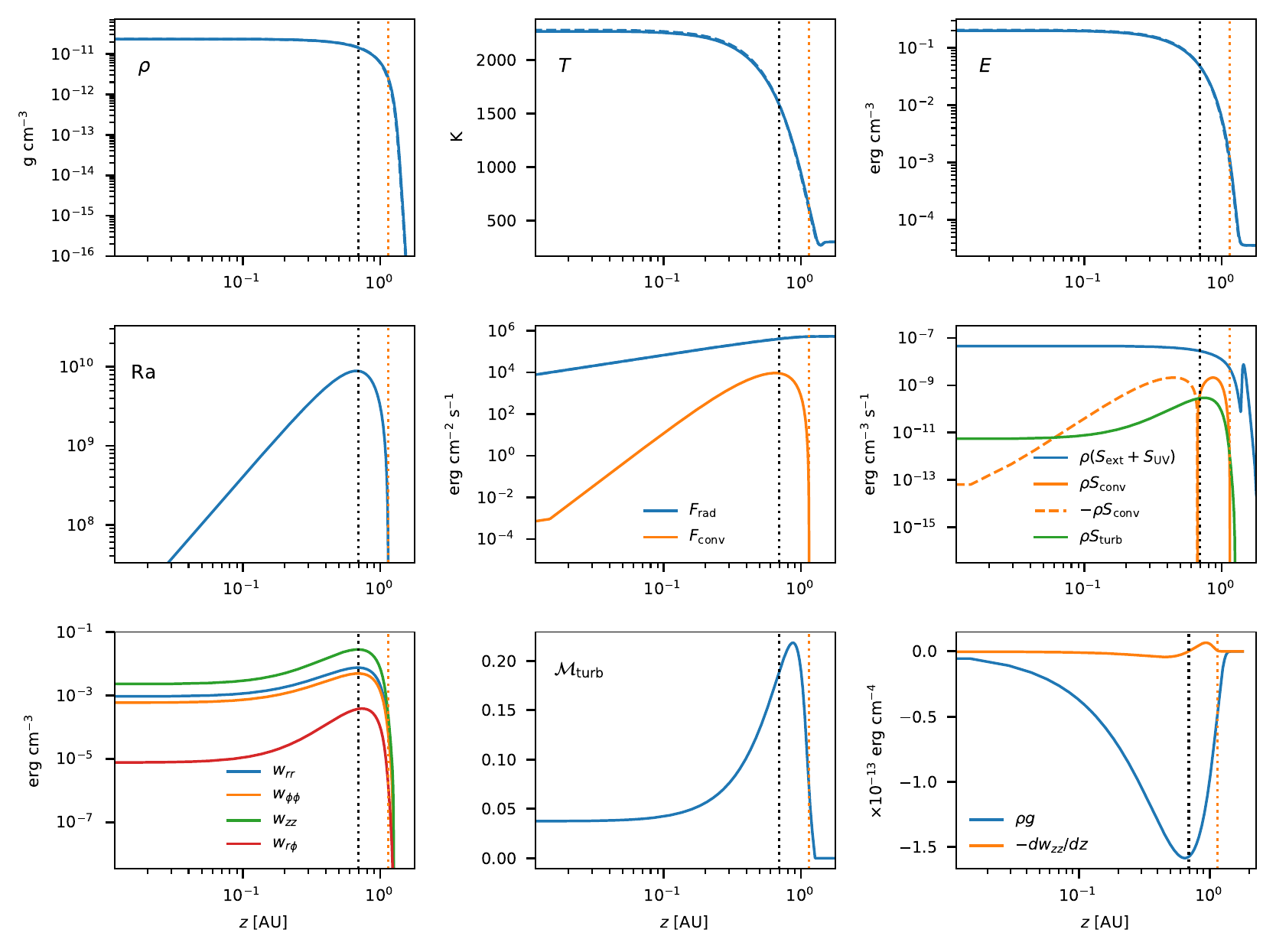}
  \end{center}
  \caption{The Model B results, $\dot{M} = 10^{-4}$ $M_\odot\:$yr$^{-1}$. Line designations are the same as in Fig.~\ref{fig:model-a}}
  \label{fig:model-b}
\end{figure*}

The Model B results are shown in Fig. \ref{fig:model-b}. Steady state was reached in eight Keplerian orbits, i.e. much faster than {in} the Model A. This is due to higher temperature of the gas in the Model B ($\sim 2000$ K) comparing to the Model A ($\sim 200$ K), hence the shorter thermal time scale. {The} radiative time scale may be estimated using a well known approximation derived for an optically thick layer, see e.g. \cite{Wu2021ApJ...923..123W}:
\begin{equation}
  t_\mathrm{rad}
  \sim \frac{c_\mathrm{v}}{c a_r}\,\frac{\kappa_\mathrm{R}}{T^3}\,\Sigma^2  \;.
\end{equation}
Since $\kappa_\mathrm{R} \propto T$ \citep{Pavlyuchenkov2020ARep...64....1P}, an order of magnitude increase in temperature results in two order of magnitude reduction of the characteristic radiation time.

In case of higher heating power $Q_\mathrm{ext}$, which corresponds to the accretion rate $\dot{M} = 10^{-4}$ $M_\odot\:$yr$^{-1}$, the convection is also developed. The magnitude of the convective flow in the Model B is much higher than in the Model A, but is much lower than the radiative flow and does not affect the thermal structure of the layer. The maximum amplitude of turbulent fluctuations, which is reached near the upper boundary of the vertical thermal scale, exceeds $0.2\:c_\mathrm{s}$. However, the $w_{r\phi}$ tensor component responsible for the transfer of angular momentum is two orders of magnitude lower than the $w_{zz}$ component, as in the quiescent model. In the Model A $\dot{M}_\mathrm{eff}/\dot{M} = 1.7\times10^{-3}$, while in the Model B $\dot{M}_\mathrm{eff}/\dot{M} = 1.7\times10^{-4}$.

\begin{table}
  \centering
  \caption{Main results of Models A and B. $\dot{M}$ is the accretion rate {parameterising} the external heat source $Q_\mathrm{ext}$; $\dot{M}_\mathrm{eff}$ is the effective accretion rate; $Q_\mathrm{UV}$ is stellar UV radiation source; $Q_\mathrm{conv}$ is the source associated with the convection; $Q_\mathrm{turb}$ is the turbulent energy dissipation rate; $W_{ij}$ are the $z$-integrated components of the Reynolds stress tensor; $\alpha_\mathrm{eff}$ is the effective Shakura-Sunyaev parameter.}
  \begin{tabular}{rcll}
    \hline  \\[-3mm]
    Quantity & Units & Model A & Model B  \\
    \hline  \\[-3mm]
    $\dot{M}$              & $M_\odot\:$yr$^{-1}$ & $10^{-7}$ & $10^{-4}$  \\
    $\dot{M}_\mathrm{eff}$ & $M_\odot\:$yr$^{-1}$ & $1.7\times10^{-10}$ & $1.7\times10^{-8}$  \\
    $Q_\mathrm{ext}$  & erg$\:$s$^{-1}\:$cm$^{-2}$ & $1.1\times10^3$ & $1.1\times10^6$  \\
    $Q_\mathrm{UV}$   & erg$\:$s$^{-1}\:$cm$^{-2}$ & $1.8\times10^4$ & $1.8\times10^4$  \\
    $Q_\mathrm{conv}$ & erg$\:$s$^{-1}\:$cm$^{-2}$ & $-4.3\times10^1$ & $-3.9\times10^3$  \\
    $Q_\mathrm{turb}$ & erg$\:$s$^{-1}\:$cm$^{-2}$ & $4.0\times10^1$ & $4.2\times10^3$  \\
    $\dot{M}_\mathrm{eff}/\dot{M}$   & & $1.7\times10^{-3}$ & $1.7\times10^{-4}$  \\
    $Q_\mathrm{turb}/Q_\mathrm{ext}$ & & $3.7\times10^{-2}$ & $3.9\times10^{-3}$  \\
    $W_{r\phi}/(\sum_i W_{ii})$      & & $3.5\times10^{-3}$ & $7.9\times10^{-3}$  \\
    $\alpha_\mathrm{eff}$            & & $1.6\times10^{-5}$ & $1.5\times10^{-4}$  \\
    \hline
  \end{tabular}
  \label{tbl:output}
\end{table}

\section{Discussion}
\label{sec:discussion}

In this paper we have used the idea that convection in protoplanetary discs is turbulent. In general, this is not necessarily the case. In order to quantify the transition of the convective flow between the laminar and turbulent regimes, let us use an empirical criterion based on the Reynolds number $\mathsf{Re} = V L /\nu_\mathrm{mol}$, where $V$ is the flow velocity, $L$ is its characteristic scale, and $\nu_\mathrm{mol}$ is the molecular kinematic viscosity coefficient. When $\mathsf{Re}$ exceeds a critical value $\mathsf{Re}_\mathrm{cr}$, the flow becomes turbulent. Different values of the critical Reynolds number correspond to different types of flow: from tens for rotational flows to $10^3$ and even $10^5$ in the special cases of the flow in a tube \citep{Landau1959flme.book.....L}. We can apply these empirical considerations to the parameters of the convective flow. Let us determine $V$ as the convective element velocity $\omega \ell$, Eq.~\eqref{eq:convective-velocity}, and $L$ to be equal to the mixing length $\ell$. After substitution of the weak convection limit for $\omega$, see Eq.~\eqref{eq:convective-increment}, the Reynolds number becomes
\begin{equation}
  \mathsf{Re}
  = \frac{|g| \beta \ell^4}{T \nu_\mathrm{mol} \nu_\mathrm{rad}}  \;.
\end{equation}
This expression is actually the Rayleigh number $\mathsf{Ra}$ \citep{Canuto1992ApJ...392..218C, Held2021MNRAS.504.2940H}. We may therefore speculate that for the laminar-turbulent transition of the convective flow, the critical Rayleigh number should be of the order of the critical Reynolds number. Typically we had $\mathsf{Ra} \gtrsim 10^9$ in our calculations, see Figures \ref{fig:model-a} and \ref{fig:model-b}, which is high enough for turbulence to develop.

The efficiency of convection in driving accretion has long been debated. At one point, a number of papers appeared which argued that the convection that develops in an accretion disc could not lead to an outward transfer of angular momentum. \cite{Stone1996ApJ...464..364S} proposed some analytical arguments to support the idea that in the accretion disc with convective turbulence the angular momentum flux is directed inward. Their argument was based on the assumption that angular variations of pressure in axisymmetric turbulent flows are subdominant compared to the radial variations (see their comments below the Eq.~(11)). This is not the case in our model, since the radial and angular components of the isotropisation tensor $\Pi_{ij}$ are comparable in magnitude. Previously, \cite{Held2021MNRAS.504.2940H} noted that in the high-resolution numerical model the hydrodynamic convection can transport angular momentum outwards \citep{Lesur2010MNRAS.404L..64L, Held2018MNRAS.480.4797H}.

The models based on the mean-field approach inevitably depend on free parameters. Given the conventional, experiment-based values of the parameters, Eq.~\eqref{eq:constants}, the angular momentum in our model is transferred outwards. The numerical experiments with a local model (Sec.~\ref{sec:testing}) have shown that the direction of the angular momentum transport depends crucially on the value of the parameter $\mathbb{C}_{\Pi 2}$. The assumed value of $\mathbb{C}_{\Pi 2}$ is quite close to a critical value, below which the angular momentum flux changes its sign.

Our motivation for studying convection in protoplanetary discs was particularly inspired by the idea that turbulent convection could arrange irregular accretion from the disc to a star, as proposed by \cite{Pavlyuchenkov2020ARep...64....1P, Maksimova2020ARep...64..815M}. The importance of convection has also been mentioned in other studies, e.g. by \cite{Hirose2015MNRAS.448.3105H, Held2021MNRAS.504.2940H}, when considering high accretion states of MRI active protoplanetary discs. However, our results indicate that turbulent convection is a weak mechanism for angular momentum transfer in the protoplanetary disc. In fact, we support the results of \cite{Lesur2010MNRAS.404L..64L} and \cite{Held2018MNRAS.480.4797H} that the turbulence generated by convection does not provide the observed disc accretion rates and sufficient heat influx for convection to be self-sustaining. There are two reasons for this: the anisotropy of the turbulence, and the fact that convection is too weak a source of turbulence.

The first reason is that the $w_{zz}$ element of the isotropic part of the Reynolds tensor is the only element excited by convection, while $w_{r\phi}$ is the only element responsible for the removal of angular momentum from the disc. The energy exchange between Reynolds stress tensor elements is not efficient enough, so $w_{r\phi}$ is more than two orders of magnitude smaller than $w_{zz}$.

To estimate the importance of the second factor, we can look at the turbulence in the steady state near the maximum of $w_{ij}$ in Fig.~\ref{fig:model-a} or \ref{fig:model-b} (bottom row, left). Under these conditions, the diffusion term disappears from the equations \eqref{eq:flow-reynolds-cyl-wrr}--\eqref{eq:flow-reynolds-cyl-wrf}, while the system of linear algebraic equations remains. The only inhomogeneous term in this system is $B_{zz}$. Thus, the solution of the resulting system is proportional to the value of this convective source, see Section~\ref{sec:testing}. To make the effective accretion rate $\dot{M}_\mathrm{eff}$ formally equal to the given $\dot{M}$, it is necessary to increase the convective flux by three orders of magnitude in the Model A and by four orders of magnitude in the Model B. This is hardly possible, even considering the uncertainty of the mixing length $\ell$.

We note that convection might still play an important role in facilitating angular momentum transport if turbulence is excited not by convection alone, but by the collective effects of different instabilities. For example, the papers by \cite{Hirose2014ApJ...787....1H, Coleman2018ApJ...857...52C, Scepi2018A&A...609A..77S} and \cite{Held2021MNRAS.504.2940H} present calculations of 3D MHD models showing that the joint action of convection and magneto-rotational instability can increase $\alpha$ to the observed values.

It would be interesting to apply our mean field model to other instabilities, such as vertical shear instability, streaming instability, etc. \citep[see the reviews by][]{Bae2022arXiv221013314B, Lesur2022arXiv220309821L}. For example, \cite{Stoll2017A&A...599L...6S} used numerical simulations to show that vertical shear instability leads to the appearance of anisotropic turbulence.

\section{Conclusions}
\label{sec:conclusions}

In this study, we have presented a model for the transport of anisotropic turbulence in a protoplanetary disc. The model includes time-dependent heat transfer by radiative diffusion and convection, developing on the background of hydrostatic equilibrium. The time-dependent turbulent transport model is based on the mean-field approach formulated in terms of Reynolds stresses. The seed of turbulence in our model is the convective instability, hence the convection flux. In addition, the turbulence interacts with the background shear flow, which  increases the amplitude and anisotropy of the turbulence. The advantage of this model is that it allows to explicitly measure the contribution of different factors involved to the cycle of thermal and turbulent energy (Fig.~\ref{fig:scheme}). At the same time, it should be noted that this approach does not allow the study of the detailed spatial and temporal structure of turbulence, but only its effect on the mean flow.

The aforementioned model was used to study turbulence driven by convection in accretion discs. Two models of protoplanetary discs have been examined, one for the quiescent state and one for the outburst state. We can agree with the results presented in \cite{Held2018MNRAS.480.4797H} that convection-induced turbulence results in the outward transfer of angular momentum. The amplitude of the turbulence (Mach number $\sim 0.1$) is in agreement with the estimates from molecular line observations \citep{Flaherty2017ApJ...843..150F, Flaherty2018ApJ...856..117F}. However, the turbulence is found to be too weak to either reproduce the heat flux through the dissipation channel ($Q_\mathrm{turb}/Q_\mathrm{ext} = 3.7\times10^{-2}$ in Model A) to support self-sustaining convection, or to generate sufficient torque to drive the accretion rate and the associated heating source ($\dot{M}_\mathrm{eff}/\dot{M} = 1.7\times10^{-3}$ in Model A). A possible explanation for this is that convection only excites the vertical component of the Reynolds stress tensor, $w_{zz}$, directly, while angular momentum transfer is controlled by the mixed component $w_{r\phi}$. The latter is weak because of weak coupling of the turbulence and background velocity shear. It would be worth investigating instabilities occurring in both the radial and azimuthal directions as a potential source of mixed component excitation. For example, subcritical baroclinic instability, magneto-rotational instability and others.

\section{Acknowledgements}

The authors are grateful to the reviewer for valuable comments and suggestions for improving the article. We also thank A.V. Tutukov and D.V. Bisikalo for helpful discussions. We are grateful to T.S. Molyarova for her help in preparing the manuscript.

The study was supported by a grant from the Russian Science Foundation No. 22-72-10029.

\section{Data Availability}

Software code and data are available at \textsc{Github} repository via \url{https://github.com/evgenykurbatov/kp23-turb-conv-ppd}




\bsp  
\label{lastpage}
\end{document}